\documentclass[12pt, preprint]{aastex}
\usepackage{bm}

\shorttitle{LEECH GJ 504 b}
\shortauthors{Skemer et al.}

\begin{document}

\title{The LEECH Exoplanet Imaging Survey: Characterization of the Coldest Directly Imaged Exoplanet, GJ 504 b, and Evidence for Super-Stellar Metallicity\footnote{The LBT is an international collaboration among institutions in the United States, Italy and Germany. LBT Corporation partners are: The University of Arizona on behalf of the Arizona university system; Istituto Nazionale di Astrophisica, Italy; LBT Beteiligungsgesellschaft, Germany, representing the Max-Planck Society, the Astrophysical Institute Potsdam, and Heidelberg University; The Ohio State University, and The Research Corporation, on behalf of The University of Notre Dame, University of Minnesota and University of Virginia.}}
\author{
Andrew J. Skemer$^{1,2}$,
Caroline V. Morley$^{2}$,
Neil T. Zimmerman$^{3,4}$,
Michael F. Skrutskie$^{5}$,
Jarron Leisenring$^{1}$,
Esther Buenzli$^{3}$,
Mickael Bonnefoy$^{3,6}$,
Vanessa Bailey$^{1}$,
Philip Hinz$^{1}$,
Denis Defr\'ere$^{1}$,
Simone Esposito$^{7}$,
D\'aniel Apai$^{1,8}$,
Beth Biller$^{3,9}$,
Wolfgang Brandner$^{3}$,
Laird Close$^{1}$,
Justin R. Crepp$^{10}$,
Robert J. De Rosa$^{11,12}$,
Silvano Desidera$^{13}$,
Josh Eisner$^{1}$,
Jonathan Fortney$^{2}$,
Richard Freedman$^{14,15}$,
Thomas Henning$^{3}$,
Karl-Heinz Hofmann$^{16}$,
Taisiya Kopytova$^{3}$,
Roxana Lupu$^{15}$,
Anne-Lise Maire$^{13}$,
Jared R. Males$^{1}$,
Mark Marley$^{15}$,
Katie Morzinski$^{1}$,
Apurva Oza$^{5,17}$,
Jenny Patience$^{11}$,
Abhijith Rajan$^{11}$,
George Rieke$^{1}$,
Dieter Schertl$^{16}$,
Joshua Schlieder$^{3,15}$,
Jordan Stone$^{1}$,
Kate Su$^{1}$,
Amali Vaz$^{1}$,
Channon Visscher$^{18}$,
Kimberly Ward-Duong$^{11}$,
Gerd Weigelt$^{16}$,
Charles E. Woodward$^{19}$
}

\affil{$^{1}$Steward Observatory, University of Arizona, Tucson, AZ, USA}
\affil{$^{2}$University of California, Santa Cruz, Santa Cruz, CA, USA}
\affil{$^{3}$Max Planck Institute for Astronomy, Heidelberg, Germany}
\affil{$^{4}$Princeton University, Princeton, NJ, USA}
\affil{$^{5}$University of Virginia, Charlottesville, VA, USA}
\affil{$^{6}$Institut de Planetologie et d'Astrophysique de Grenoble, Grenoble, France}
\affil{$^{7}$Istituto Nazionale di Astrofisica - Arcetri Astrophysical Observatory, Florence, Italy}
\affil{$^{8}$Lunar and Planetary Laboratory, University of Arizona, Tucson, AZ, USA}
\affil{$^{9}$University of Edinburgh, Edinburgh, UK}
\affil{$^{10}$Notre Dame University, South Bend, IN, USA}
\affil{$^{11}$Arizona State University, Tempe, AZ, USA}
\affil{$^{12}$University of California, Berkeley, Berkeley, CA, USA}
\affil{$^{13}$Istituto Nazionale di Astrofisica - Padova Astronomical Observatory, Italy}
\affil{$^{14}$Search for Extraterrestrial Intelligence Institute, Mountain View, CA, USA}
\affil{$^{15}$NASA Ames Research Center, Moffett Field, CA, USA}
\affil{$^{16}$Max Planck Institute for Radio Astronomy, Bonn, Germany}
\affil{$^{17}$Universit\'e Pierre et Marie Curie, Paris, France}
\affil{$^{18}$Dordt College, Sioux Center, IA, USA}
\affil{$^{19}$Minnesota Institute for Astrophysics, University of Minnesota, Minneapolis, MN, USA}

\begin{abstract}
As gas giant planets and brown dwarfs radiate away the residual heat from their formation, they cool through a spectral type transition from L to T, which encompasses the dissipation of cloud opacity and the appearance of strong methane absorption. While there are hundreds of known T-type brown dwarfs, the first generation of directly-imaged exoplanets were all L-type.  Recently, Kuzuhara et al. (2013) announced the discovery of GJ 504 b, the first T dwarf exoplanet.  GJ 504 b provides a unique opportunity to study the atmosphere of a new type of exoplanet with a $\sim$500 K temperature that bridges the gap between the first directly imaged planets ($\sim$1000 K) and our own Solar System's Jupiter ($\sim$130 K). We observed GJ 504 b in three narrow L-band filters (3.71, 3.88, and 4.00 $\micron$), spanning the red end of the broad methane fundamental absorption feature (3.3~\micron) as part of the LEECH exoplanet imaging survey.  By comparing our new photometry and literature photometry to a grid of custom model atmospheres, we were able to fit GJ 504 b's unusual spectral energy distribution for the first time.  We find that GJ 504 b is well-fit by models with the following parameters: T$_{eff}$=544$\pm$10 K, g$<$600 $m/s^{2}$, [M/H]=0.60$\pm$0.12, cloud opacity parameter of f$_{sed}=2-5$, R=0.96$\pm$0.07 R$_{Jup}$, and log(L)=-6.13$\pm$0.03 L$_{\Sun}$, implying a hot start mass of 3-30 M$_{jup}$ for a conservative age range of 0.1-6.5 Gyr.  Of particular interest, our model fits suggest that GJ 504 b has a super-stellar metallicity.  Since planet formation can create objects with non-stellar metallicities, while binary star formation cannot, this result suggests that GJ 504 b formed like a planet, not like a binary companion.
\end{abstract}

\section{Introduction}  
When brown dwarfs cool below $\sim$1,200 K, their atmospheres transition from cloudy to clear, and methane becomes a dominant absorber in their spectral energy distributions.  Although gas giants were thought to be analogs to brown dwarfs, the first generation of directly imaged exoplanets had cloudy, methane-free atmospheres, even though their temperatures are well below the temperature where field brown dwarfs have had their ``L$\rightarrow$T'' transition \citep{2004AA...425L..29C,2011ApJ...732..107S,2008Sci...322.1348M,2010Natur.468.1080M}.  GJ 504 b, discovered by the SEEDS survey \citep[Strategic Explorations of Exoplanets and Disks with Subaru;][]{2009AIPC.1158...11T}, is the first example of an exoplanet that is cold enough ($\sim$500K) to be relatively cloud-free and have strong methane absorption features \citep{2013ApJ...774...11K,2013ApJ...778L...4J}.  Another T-dwarf exoplanet was recently discovered around 51 Eri \citep{2015Sci...350...64M}.

At a separation of 2\farcs5 (43.5 AU) from its G-star host, GJ 504 b is easily accessible to most high-contrast imaging systems \citep{2013ApJ...774...11K}.  Its H-Ks color (0.63$\pm$0.15) is highly discrepant with similar luminosity field brown dwarfs \citep[$\sim$-0.2;][]{2012ApJS..201...19D}. At the same time it has strong methane absorption at 1.66$\micron$, clearly placing it in a different class than other directly imaged exoplanets \citep{2013ApJ...778L...4J}.  GJ 504 also has super-solar metallicity \citep[{[}M/H{]}=0.10-0.28, although most determinations are toward the lower end of this range;][]{1993A&A...275..101E,2004A&A...418..551M,2005ApJS..159..141V,2007PASJ...59..335T,2010MNRAS.403.1368G,2012A&A...541A..40M,2013ApJ...764...78R}.  It is therefore relatively likely to have a gas giant planet \citep{2005ApJ...622.1102F}.

The age of GJ 504 A, and thus the mass and planetary status of GJ 504 b, is uncertain. \citet{2013ApJ...774...11K} find consistency among multiple age indicators, such as X-ray activity, rotation rate, chromospheric activity, and HR diagram location, indicating an age for GJ 504A of 0.1-0.5 Gyr. However, a reanalysis of GJ 504 A's stellar properties by \citet{2015ApJ...806..163F} suggests that the star lies above the main sequence on an HR diagram with a corresponding age of ~4.5 Gyr. \citet{2015ApJ...806..163F} argue that the rapid rotation and other signs of youth arise because a massive planet has fallen into the star, carrying its orbital angular momentum with it. This leaves the presence of strong lithium absorption \citep{2013ApJ...774...11K} unexplained \citep{2010IAUS..268..359S}. Some of the lithium could have been replenished by the planet \citep{2012ApJ...757..109C}, but an usually massive planet would be required.  Since there is no consensus on the age of the system, we consider both estimates in the following discussion; the younger age range implies a planet mass of $\sim$ 3-9 Mjup, while the older one would suggest a mass of $\sim$30 Mjup. 

The LBTI Exozodi Exoplanet Common Hunt (LEECH) is a $\sim$100 night survey with the Large Binocular Telescope (LBT) to search for and characterize exoplanets in the mid-infrared \citep{2014SPIE.9148E..0LS,2015A&A...576A.133M}.  In this work, we confirm the detection of GJ 504 b, and obtain photometry of the first T-dwarf exoplanet in 3 narrow L-band filters (3.71, 3.88, and 4.00 $\micron$).  For T-dwarfs, L-band photometry can probe the broad methane fundamental absorption feature (centered at 3.3~\micron) and put strong constraints on the luminosity of the planet, which peaks at $\sim$4$\micron$.  The overall benefit of this additional photometry is to improve our ability to constrain GJ 504 b's bulk properties with atmospheric modeling.  In Section 2, we present our observations and reductions, which comprise some of the deepest images taken from the ground at these wavelengths.  In Section 3, we present our new photometry, and adjust the literature photometry onto a common photometric system. In Section 4, we fit the photometry with a grid of models and discuss the physical nature of GJ 504 b.  Finally we present our conclusions in Section 5.

\section{Observations and Reductions}
We observed GJ 504 on UT April 21, 2013 and UT March 11-13, 2014 with the Large Binocular Telescope Interferometer \citep[LBTI; ][]{2012SPIE.8445E..0UH} and its 1-5 $\micron$ imager, L/M Infrared Camera \citep[LMIRcam;][]{2010SPIE.7735E.118S,2012SPIE.8446E..4FL}.  The LBT has twin deformable secondary adaptive optics (AO) systems \citep{2011SPIE.8149E...1E,2014SPIE.9148E..03B}, which provide excellent sensitivity in the thermal infrared ($\gtrsim$2$\micron$) compared to traditional AO systems \citep{2000PASP..112..264L}.  The diffraction-limited beams from the AO systems are fed into LBTI, which can overlap or separate the two images on LMIRcam.  For contrast-limited observations, such as the LEECH planet search \citep{2014SPIE.9148E..0LS}, we typically separate the beams to allow independent and redundant observations of the inner, speckle-noise limited regime.  For sensitivity-limited observations, we overlap the beams (incoherently) so that the faint astronomical source can be extracted from a single bright sky background aperture, rather than two.  GJ 504 b, at a separation of $\sim$2~\farcs5, falls into the latter category, where sensitivity is a greater priority than contrast.  For the UT April 21, 2013 observations, the LBT's right-side adaptive optics system was unavailable so we acquired data using just the left-side of the telescope.  For the UT March 11-13, 2014 observations, both sides of the telescope were operable, and we overlapped the two images of GJ 504 b to increase our sensitivity.

\begin{figure}
\begin{center}
\includegraphics[angle=90,width=\columnwidth]{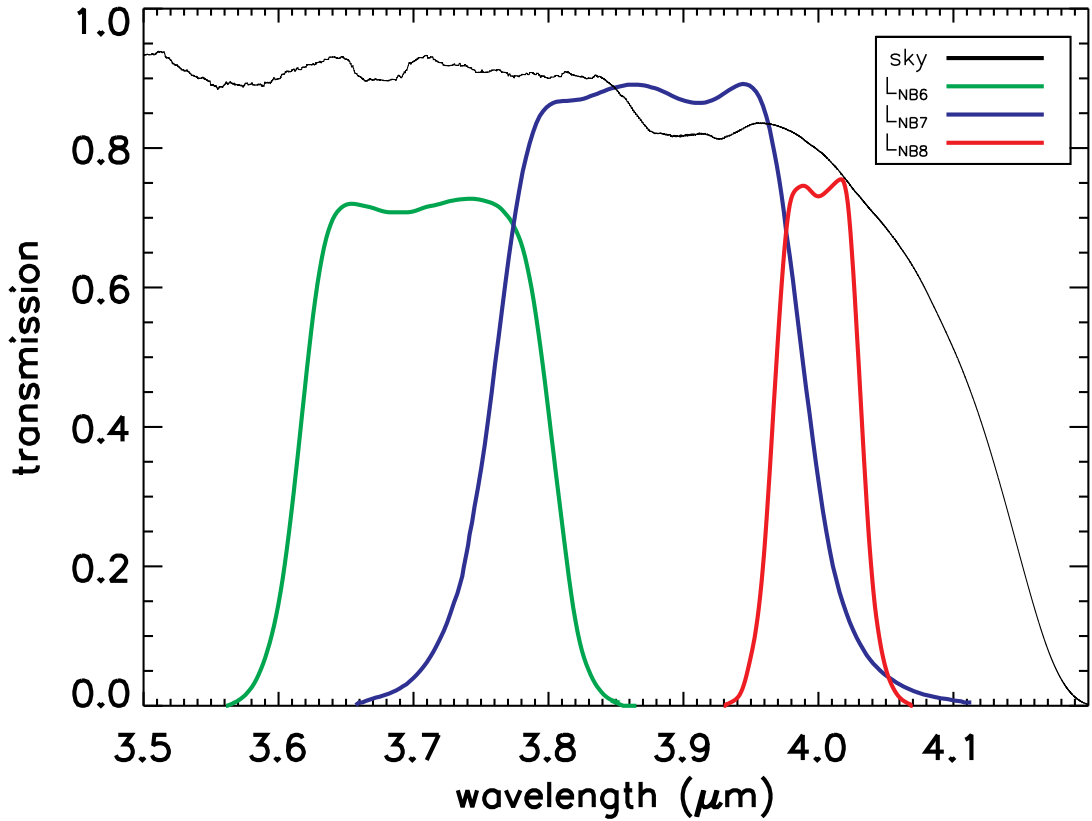}
\caption{Transmission profiles for the filters used in this paper and a telluric transmission profile for 1.0 airmasses and 4.3 mm precipitable water vapor from Gemini (http://www.gemini.edu/sciops/telescopes-and-sites/observing-condition-constraints/ir-transmission-spectra).
\label{filter profiles}}
\end{center}
\end{figure}

We observed GJ 504 in narrow-band filters: L$_{NB6}$ (3.61-3.80$\micron$), L$_{NB7}$ (3.76-3.99$\micron$), and L$_{NB8}$ (3.97-4.03$\micron$).  Basic properties for these filters are tabulated in Table \ref{filter table} and transmission profiles are shown in Figure \ref{filter profiles}.  The weather was photometric on the nights we obtained L$_{NB6}$ (UT, March 12, 2014) and L$_{NB7}$ (UT April 21, 2013) data.  The first night of L$_{NB8}$ observations (UT 2014 March 11) was non-photometric, which prompted us to repeat this filter on UT 2014 March 13.  UT 2014 March 13 had patchy clouds away from the telescope, which cleared early in the observations.  Integration times were chosen to be long enough that the off-star data were sky background-noise limited, rather than read-noise limited.  This choice meant that the star was saturated in the frames used to detect GJ 504 b.  Additional frames with shorter integration times were used to measure the brightness of the host star.  A summary of our observations is presented in Table \ref{observations}.

We reduced the data using a custom LMIRcam pipeline developed at MPIA \citep{2014AA...562A.111B}.  The pipeline (1) replaces bad detector pixels with the median of their adjacent neighbors, based on a table of outlier pixels cataloged from off-sky calibration frames, (2) removes the detector bias and background sky/telescope emission by subtracting the median of images from chronologically neighboring nod sub-sequences, (3) determines the sub-pixel centroid of the star point-spread-function (PSF) in each image by fitting a Gaussian, masking the inner saturated pixels, and shifts and crops the image to a common, aligned frame, (4) flags images with peak star fluxes below a specified threshold and images with abnormal background levels to exclude data contaminated by clouds and poor seeing, (5) removes residual detector bias from columns and rows, based on overscan pixels, (6) forms the cube of reduced, photometric quality images, for inspection and PSF subtraction.  Angular differential imaging (ADI) and principal component analysis (PCA) PSF subtraction is carried out with a separate program described later in this section.

\begin{figure}
\begin{center}
\includegraphics[width=0.9\textwidth]{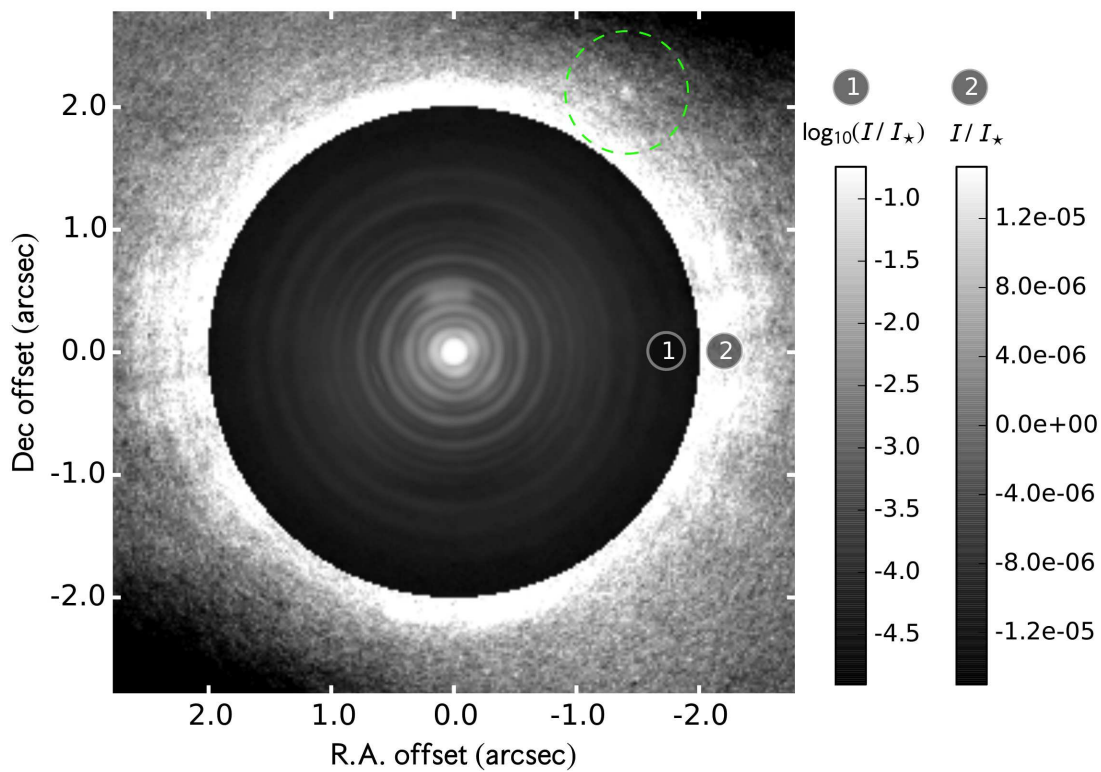}
\caption{De-rotated co-add of all photometric quality GJ 504 images taken in the L$_{NB7}$ (3.95 \micron) filter. The planet, GJ 504 b, is the circled point source to the northwest of the star, at separation $\sim$2~\farcs5. Due to the large dynamic range of structure in the image, two regions are displayed with different flux scales: (1) within 2\arcsec~of the star, the Airy rings are displayed with a logarithmic stretch; the outer region (2) that includes the planet is displayed with a linear stretch centered at zero. The corresponding grayscale bars on the right-hand side indicate the flux in units normalized to the peak of the unsaturated star's PSF.
\label{two-stretch image}}
\end{center}
\end{figure}

Figure \ref{two-stretch image} shows the result of aligning de-rotating, and co-adding all of the photometric quality images acquired of GJ 504 b in the $L_{\rm NB7}$ (3.95~\micron) filter -- a composite with an effective integration time of 3121 s. Owing to the relatively wide angular separation of GJ 504 b, the planet is visible without subtracting the star, in the northwest corner of the co-added image. The image flux scale is given in units of the star's peak intensity, determined from the short exposure (unsaturated PSF) image sequence.

\begin{figure}
\begin{center}
\includegraphics[width=0.9\textwidth]{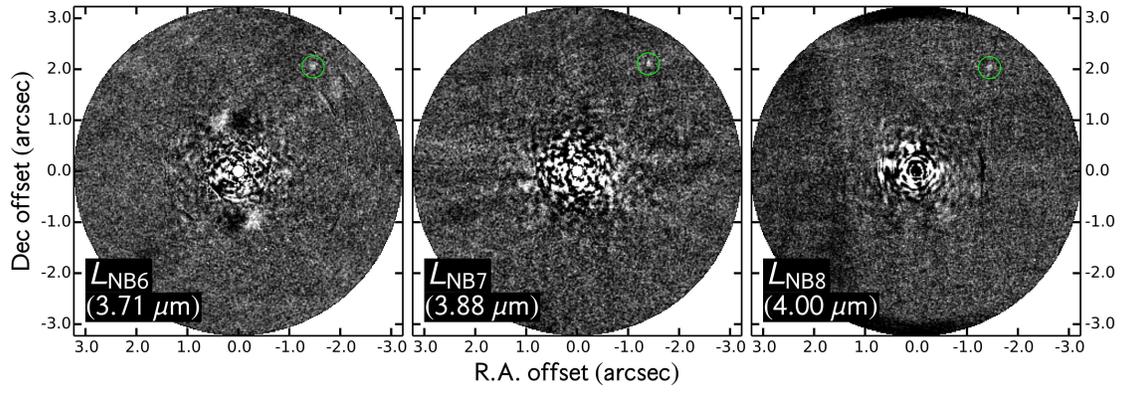}
\caption{Final starlight-subtracted images of data taken in the L$_{NB6}$, L$_{NB7}$, and L$_{NB8}$ filters.
\label{PSF-subtracted images}}
\end{center}
\end{figure}

To reduce starlight contamination in the planet signal, we subtract an estimate of the star's PSF from each image using standard high-contrast imaging techniques.  The data were taken in angular differential imaging mode \citep[ADI;][]{2006ApJ...641..556M}, where the instrument does not rotate (LBTI is on a non-rotating mount) so that its diffractive pattern and static aberrations stay fixed, while the sky image rotates with parallactic angle.  We then reduced the data with a PCA high-contrast algorithm \citep{2012ApJ...755L..28S,2012MNRAS.427..948A,2014ApJ...794..161F}, using a custom implementation that closely follows \citet{2012ApJ...755L..28S}.  At each separation, we optimize the number of principle components that are used in the subtraction ($N_{PC}$) by inserting 12 artificial planets at different azimuth angles, and measuring their signal-to-noise ratios as a function of $N_{PC}$.  We also optimize the minimum parallactic rotation gap (parameterized as $N_{FHWM}$) between the image being fit, and the library of images used to do the fit \citep{2007ApJ...660..770L}.  Optimal values of $N_{PC}$ vary from $\sim$15-30, and optimal values of $N_{FHWM}$ vary from $\sim$0.5-1.0.  Final reduced images are shown in Figure \ref{PSF-subtracted images}.  We detect GJ 504 b at signal-to-noise of 5.9, 5.3, and 5.9 in $L_{NB6}$, $L_{NB7}$, and $L_{NB8}$ respectively (as described below).


Relative photometry between GJ 504 A and b is measured using the forward modeling approach described in \citet{2012ApJ...755L..28S}, which uses an image of the star at the position of the planet to calibrate the planet self-subtraction that is common to high-contrast image processing techniques \citep{2007ApJ...660..770L}.  We optimize the source model over an annular sector centered on the planet and spanning 30 degrees in position angle, conservatively fixing $N_{FHWM}=1.0$, as we find that values less than this increase self-subtraction.  We measure the error on our relative photometry by inserting artificial planets at the same separation as GJ 504 b but at different position angles, and repeating the forward modeling.  The standard deviation of our measurements of the artificial planets is adopted as our formal relative photometry uncertainty, and our detection signal to noise.  Relative photometry between GJ 504 A and b is reported in Table \ref{photometry}.  Note that the two nights of L$_{NB8}$ observations were combined in this analysis, after confirming that they produced similar photometry, within 1-$\sigma$ errors.

\section{Photometry}\label{photometry section}

To convert contrast measurements (relative photometry) to apparent magnitudes, we adopt and calculate photometry for GJ 504 A.  For J, H and Ks, we adopt apparent photometry from \citet{2003AJ....125.3311K}, converted to 2MASS photometry (J=4.13$\pm$0.01, H=3.88$\pm$0.01, Ks=3.81$\pm$0.01) using the methodology and Vega spectrum of \citet{2008AJ....135.2245R}.  For all other filters where GJ 504 b has been observed, we calculate GJ 504 A's apparent photometry by fitting a model stellar atmosphere \citep{2004astro.ph..5087C} with parameters that best match  GJ504's temperature, gravity and metallicity measurements \citep{2005ApJS..159..141V}.  In our fit, we use the previously quoted JHKs photometry as well as WISE \citep{2013yCat.2328....0C} W3 (3.831$\pm$0.015) and W4 (3.757$\pm$0.022) photometry (the WISE W1 and W2 filters were not used because they were flagged as saturated).  Our fit to these 5 data points with a model atmosphere produced a reduced $\chi^{2}$ of 0.70.  We estimate GJ 504 A's apparent magnitude to be 3.87 in CH$_{4}$s, 3.85 in CH$_{4}$l, and 3.82 in L', L$_{NB6}$, L$_{NB7}$, and L$_{NB8}$\footnote{Our L estimates (for all four filters) are 0.12 mag different than the L' estimate of \citet{2013ApJ...774...11K}, who used a photometric measurement of GJ 504 A with a large (0.09 mag) uncertainty.  Our model-based estimate is more precise and is consistent with the JHK photometry.}.  This model-driven approach leads to a 1-2$\sigma$ inconsistency between our CH$_{4}$ photometry and our H-band photometry (the two CH$_{4}$ bands span H-band and should not be brighter than the H photometry), which we would like to correct in order to avoid propagating erroneous color information into GJ 504 b's photometry (incorrect colors could affect our derived model atmosphere parameters, whereas an incorrect overall luminosity will only affect radius).  Therefore, we adjust our estimated CH$_{4}$ photometry to be fainter by 0.02 mag to be consistent with the broad H-band value.  Similarly, we adjust the L photometry to be brighter by 0.02 mag to be consistent with the Ks photometry.  Since inconsistencies in our photometric estimates appear to be at the $\sim$0.02 mag level, we adopt 0.02 mag uncertainties for all GJ 504 A photometry.  These inconsistencies are independent of our exact choice of stellar parameters, and are more likely the result of propagated photometric errors or stellar variability. Our adopted GJ 504 A photometry along with the resulting GJ 504 b photometry are summarized in Table \ref{photometry}.  For our atmosphere modeling, we convert to absolute magnitudes assuming a distance of 17.56$\pm$0.08 pc \citep{2007AA...474..653V}.

\section{Discussion: GJ 504 b's Unusual Appearance}
With the first discoveries of directly imaged planets, it was obvious that something about planetary atmospheres made them different than the atmospheres of similar temperature field brown dwarfs \citep{2004AA...425L..29C,2008Sci...322.1348M}.  In near-infrared color-magnitude diagrams (see Figure \ref{cmd}), the HR 8799 planets fall on what appears to be an extension of the L-dwarf sequence.  The implication is that these young planets have retained their dusty, methane poor atmospheres at lower luminosities than old field brown dwarfs, which transition to methane-rich, cloud-free T-dwarfs below $M_{H}\sim$14 magnitudes.  Surveys have now found young, dusty brown dwarfs that are more analogous to the HR 8799 planets \citep[e.g. ][]{2013AJ....145....2F,2013ApJ...777L..20L,2015ApJ...804...96G}, and there are theoretical justifications for why, in addition to effective temperature, an object's mass/surface gravity affects the clouds and chemistry of its photosphere \citep{2012ApJ...754..135M,2014ApJ...797...41Z}.

Also plotted in Figure \ref{cmd} is GJ 504 b, which is much less luminous than previously discovered exoplanets.  In the J-H vs. J and Ks-L' vs. Ks diagrams, GJ 504 b falls right on the late T-dwarf sequence.  However, in the H-Ks vs. H diagram, GJ 504 b is much redder than the field T-dwarfs.  There is one other object near GJ 504 b in the color-magnitude diagrams: GJ 758 B, a $\sim$30-40M$_{jup}$ companion to a $\sim$6 Gyr Sun-like star \citep{2009ApJ...707L.123T,2011ApJ...728...85J,2008ApJ...687.1264M}.  

Clearly there is something about GJ 504 b and GJ 758 B that make them different than other objects with similar luminosities.  For the HR 8799 planets, youth and low surface gravity are responsible for their unusual appearances \citep{2008Sci...322.1348M}.  While GJ 504 b's age is uncertain, and GJ 758 B is clearly old, their position in Figure \ref{cmd} unambiguously demonstrates that at least one of their physical properties is unusual.  

For GJ 504 b, \citet{2013ApJ...774...11K} and \citet{2013ApJ...778L...4J} suggest that gravity or metallicity could be driving the planet's unusual near-infrared colors.  In the rest of this section, we attempt to model GJ 504 b's atmosphere and directly constrain these properties.

\subsection{Atmosphere Models}
\begin{figure}
\begin{center}
\includegraphics[angle=90,width=\columnwidth]{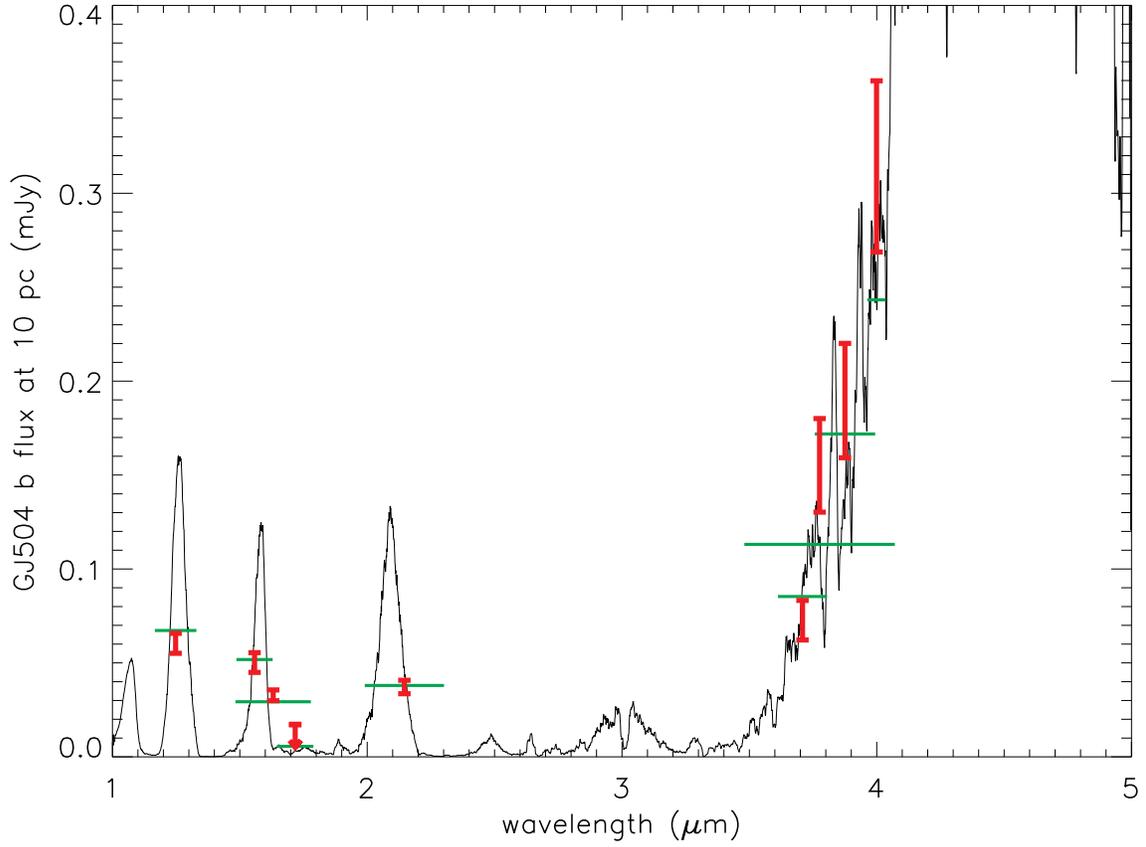}
\caption{Photometry of GJ 504 b with a best-fit model photosphere.  The red error bars and 1.7 $\micron$ upper limit are photometry tabulated in Table \ref{photometry}.  The green horizontal bars are model photometry, with the width of the bars denoting the filter bandpass.  The black curve is the best-fit from a grid of atmosphere models that vary temperature, surface gravity, metallicity, and cloud type, with radius scaling as a free-parameter.  The reduced $\chi^{2}$ (counting only radius scaling as a free-parameter) is 0.98.
\label{best-fit}}
\end{center}
\end{figure}

We attempt to fit the complete spectral energy distribution (SED) of GJ 504 b using the photometry described in Section \ref{photometry section}. We use models similar to those described in \citet{2012ApJ...756..172M,2014ApJ...787...78M}, which include opacities for T/Y-dwarf condensates.  The methane line lists have been updated using \citet{2014PNAS..111.9379Y} and the alkali line lists have been updated to use the results from \citet{2005A&A...440.1195A}.  Chemical equilibrium grids based upon previous thermochemical models \citep{1999ApJ...519..793L,2002ApJ...577..974L,2002Icar..155..393L,2006asup.book....1L,2006ApJ...648.1181V,2010ApJ...716.1060V,2012ApJ...757....5V,2013ApJ...763...25M} have been revised and extended to include higher metallicities. These updates will be described in detail in a set of upcoming papers that focus on the new model grid.  In addition to temperature, our model grid parameterizes surface gravity, metallicity, and cloud thickness\footnote{Cloud opacity is parameterized as a particle sedimentation efficiency, labeled f$_{sed}$, as described by \citet{2001ApJ...556..872A}.  Lower f$_{sed}$ numbers correspond to larger cloud opacities.}.  We allow radius to be a free parameter so that atmospheric properties rather than luminosity drive the fit.  Our model grid contains 480 models, comprising temperatures of 450, 475, 500, 525, 550, 575, 600 and 625 K, surface gravities, of 30, 100, 300 and 1000 $m/s^{2}$, metallicities of [M/H]=0, 0.5, and 1.0, and cloud thicknesses of f$_{sed}$=1, 2, 3, 5 and cloud-free.  The parameters that we choose to vary are among the most fundamental to the bulk appearance of planetary atmospheres.  However, we cannot rule out that additional parameters, such as non-equilibrium NH$_{3}$ chemistry \citep{2014ApJ...797...41Z} might play an important role as well.  We also note that systematic differences between model families can account for substantially disparate parameter estimates \citep{2012A&A...540A..85P}.  With these caveats in mind, our best-fit model is T=550K, [M/H]=0.5, R=0.94 R$_{jup}$, g=100 $m/s^{2}$, f$_{sed}$=3, and log(L)=-6.13 $L_{\Sun}$ (See Figure \ref{best-fit}).  The reduced $\chi^{2}$ (counting only radius scaling as a free-parameter) is 0.98 with 7 degrees of freedom.  None of the other models provide a fit with a relative probability higher than 1\%, based on the Bayesian probability: $P(model_{1}/model_{2})=e^{(\chi^2_{model_{2}}-\chi^2_{model_{1}})/2}$ for Gaussian photometric errors and a uniform model prior.  While this analysis demonstrates that there is at least one self-consistent model that adequately fits all of the data, our current grid is too sparse to sample the error distribution of each parameter.  Without sampling the error distribution, our best-fit model may not be at the peak of the global probability distribution.  Thus, in Section 4.2, we interpolate between the models to form a denser grid, which we use to adopt estimates for each parameter.

\begin{figure}
\begin{center}
\includegraphics[angle=90,width=\columnwidth]{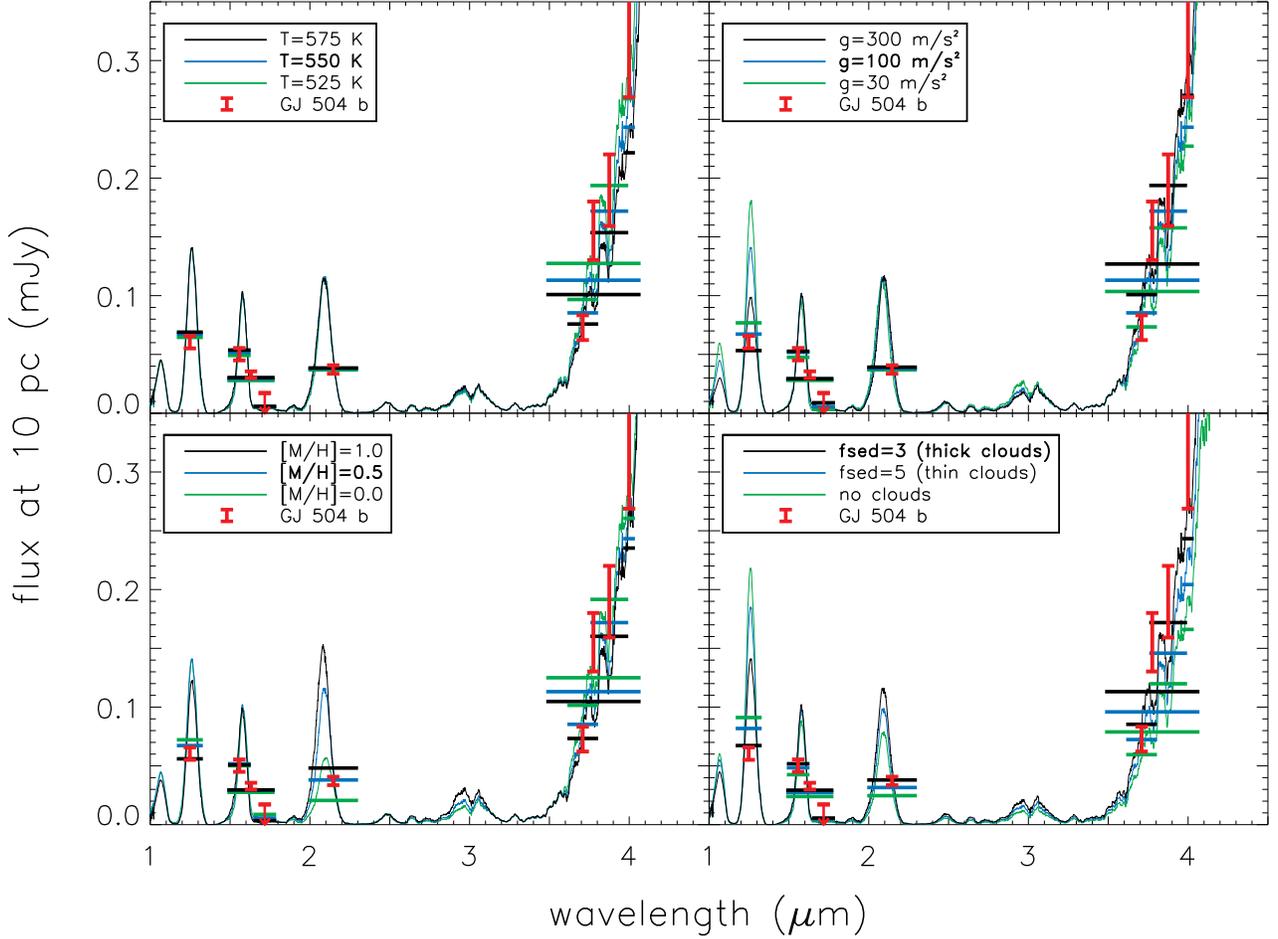}
\caption{Effects of varying model parameters.  For temperature, surface gravity, metallicity, and cloud type, we show the best-fit model (bolded in each legend) along with additional models that vary that parameter.  All models have radius scaling as a free parameter.  In this scheme (which is partially driven by the size of the error bars and the radius fit), temperature primarily affects 3-4 $\micron$ photometry.  Gravity affects the J and the 3-4 $\micron$ photometry.  Metallicity affects J, Ks and the 3-4 $\micron$ photometry.  Cloud properties also affect J, Ks and the 3-4 $\micron$ photometry.  However, for metallicity, Ks and the 3-4 $\micron$ move in opposite directions, while for cloud properties, they move in the same direction.  Thus the parameters are not degenerate.
\label{model parameters}}
\end{center}
\end{figure}

To lend some intuition to the effects of varying individual parameters, Figure \ref{model parameters} shows the best fit model in four panels, which individually vary temperature, surface gravity, metallicity, and cloud properties.  As with our best-fit, all models are scaled by radius.  In this scheme (which is partially driven by the size of the error bars and the radius fit), temperature is primarily constrained by the L' and narrow L-band photometry.  Gravity is constrained by the photometry in the J and L bands.  Metallicity is constrained by J, Ks and L.  Cloud properties are also constrained by J, Ks and L.  However, for metallicity, Ks and L move in opposite directions, while for cloud properties, they move in the same direction.  No parameter is fully degenerate with a combination of other parameters, and all four parameters (plus radius) were necessary to obtain an adequate best-fit.  The L-band photometry, in particular, was critical for resolving degeneracies between temperature and the other parameters.

\subsection{Interpolated Atmosphere Models}
\begin{figure}
\begin{center}
\includegraphics[angle=90,width=\columnwidth]{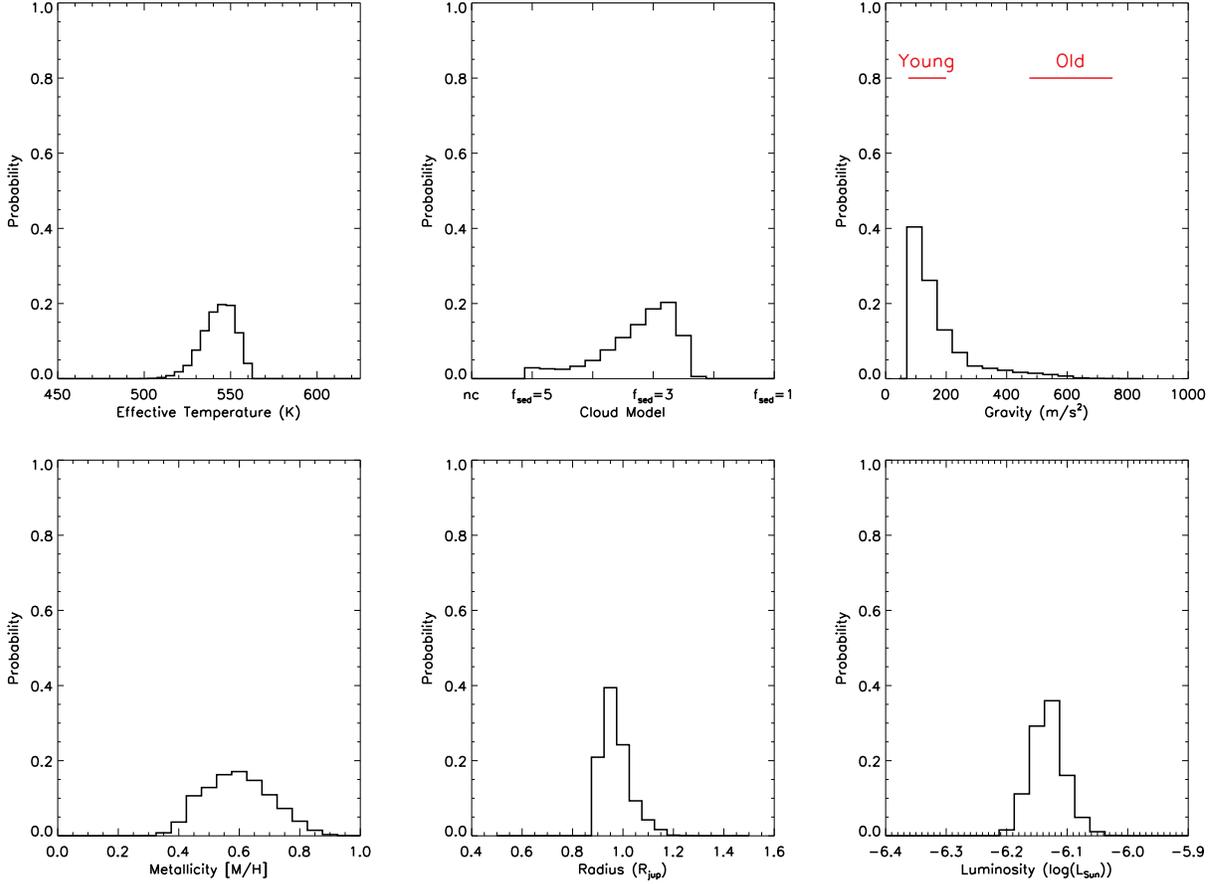}
\caption{Marginalized probability distributions of various model parameters derived by finely interpolating the model grid described in Section 4.1 (black histograms).  Fits to the marginalized probability distributions give the following marginalized parameter distributions for GJ 504 b: T$_{eff}$=544$\pm$10 K, f$_{sed}$(cloudiness)=2-5, g$<$600 $m/s^{2}$, [M/H]=0.60$\pm$0.12, R=0.96$\pm$0.07 R$_{\Sun}$, log(L)=-6.13$\pm$0.03 L$_{\Sun}$.  The red horizontal lines in the surface gravity panel correspond to the ranges consistent with the \citet{2013ApJ...774...11K} (young) and \citet{2015ApJ...806..163F} (old) age estimates, assuming luminosity-based masses from the \citet{2003A&A...402..701B} evolutionary models and R=1.0 R$_{\Sun}$.
\label{interpolated models}}
\end{center}
\end{figure}

\begin{figure}
\begin{center}
\includegraphics[angle=90,width=\columnwidth]{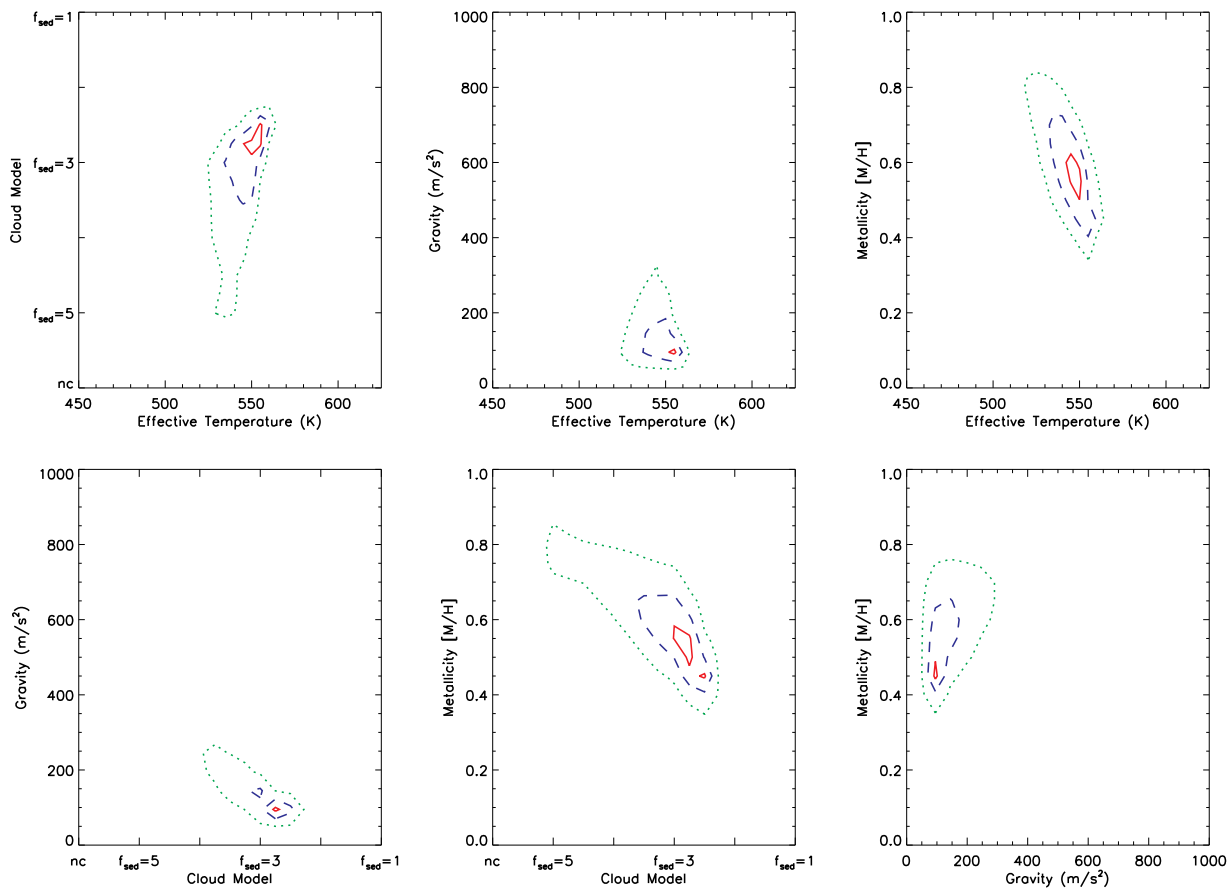}
\caption{Contour plots of probability distributions for each pair of atmosphere model parameters.  Models with 90\%, 50\% and 10\% of the peak probability are shown with red solid contours, blue dashed contours, and green dotted contours respectively.
\label{contour}}
\end{center}
\end{figure}

While our model grid is able to produce a plausible fit to the available GJ 504 b data, it is too sparsely spaced to sample the error distribution of the model parameters.  We cannot easily produce a much larger grid of models, so instead, we interpolate between the models using quad-linear interpolation (linear with temperature, metallicity and cloud properties, logarithmic with surface gravity).  We assign probabilities to each model using the Bayesian posterior, $P \propto e^{-\chi^{2}/2}$, with Bayesian priors set by grid spacing.  We adopt uniform priors for temperature, $f_{sed}$ and [M/H] over their full model-allowed range.  For gravity, we adopt uniform priors above a minimum surface gravity that is set by the radius of the planet and the minimum planet mass (3 $M_{jup}$) derived by \citet{2013ApJ...774...11K}. For radius we adopt a uniform prior between 0.9 $R_{jup}$ and 1.3 $R_{jup}$, the plausible radius range for GJ 504 b, varying mass, core mass, initial entropy, and metallicity \citep{2007ApJ...659.1661F,2008ApJ...683.1104F}.  Marginal probabilities are shown in Figure \ref{interpolated models}.  Gaussian fits to the marginalized probability distributions give the following marginalized parameter distributions for GJ 504 b: T$_{eff}$=544$\pm$10 K, [M/H]=0.60$\pm$0.12, R=0.96$\pm$0.07 R$_{\Sun}$, log(L)=-6.13$\pm$0.03 L$_{\Sun}$.  For cloudiness and gravity, whose distributions do not resemble Gaussians, we adopt f$_{sed}$(cloudiness)=2-5, and g$<$600 $m/s^{2}$.  Parallax uncertainty has a negligible effect on the radius and luminosity errors.  Two-dimensional probability contours, for each pair of model parameters, are shown in Figure \ref{contour}.  Correlations are evident between parameters.

Interpolation carries an intrinsic risk that the spectra change in non-linear ways, which can vary by bandpass.  For our particular model grid and photometry, the best example of this behavior is seen in the metallicity plot of Figure \ref{model parameters}.  The J-band photometry changes more quickly between [M/H]=0.5 and 1.0 than between [M/H]=0.0 and 0.5.  At the same time, many of the other model photometry points move linearly with [M/H].  To test how this might affect our final marginalized probability distributions, we repeated our analysis interpolating exponentially ($10^{[M/H]}$) in the J-band, and linearly in the other bands.  The net result is a shift in the probability distribution of 3K for T$_{eff}$, 0.02 for [M/H], 0.02 R$_{jup}$ for radius, and negligible changes for the other parameters.  In all cases, this shift is much smaller than our derived error bars.

\subsection{The Physical Properties of GJ 504 b}

Our model fitting constrains several physical properties of the planet that can help us understand its formation and evolution:

\subsubsection{Temperature and Radius}
We estimate GJ 504 b's temperature to be T$_{eff}$=544$\pm$10 K and its radius to be R=0.96$\pm$0.07 R$_{Jup}$.  To first order, these parameters are highly correlated (for a blackbody, $T_{eff} \propto R^{-0.5}$).  At this point, ultra-cool atmosphere models are not calibrated to the point that the radius estimate could tell us much about the planet's internal structure.

\subsubsection{Luminosity and Mass}
Our derived bolometric luminosity of log(L)=-6.13$\pm$0.03 L$_{\Sun}$ is only somewhat lower than \citet{2013ApJ...774...11K}'s estimate of -6.09$^{+0.06}_{-0.08}$.  \citet{2013ApJ...774...11K} estimate bolometric luminosity by averaging the \citet{2003A&A...402..701B} model values that correspond with their individual photometric points.  The anomalously bright Ks photometry, as they note, biases this value towards higher luminosities.  Because our atmosphere models are able to fit all of the photometry simultaneously, they provide a more reliable indicator of bolometric luminosity.  

Our revised bolometric luminosity motivates an updated estimate of GJ 504 b's mass.  Using \citet{2013ApJ...774...11K}'s age (0.1-0.51 Gyr) and \citet{2003A&A...402..701B} models, we find GJ 504 b has a mass of 3-8 M$_{jup}$.  Using \citet{2015ApJ...806..163F}'s age (3-6.5 Gyr) and \citet{2003A&A...402..701B} models, we find GJ 504 b has a mass of 19-30 M$_{jup}$.

\subsubsection{Metallicity}
With a metallicity of [M/H]=0.60$\pm$0.12, GJ 504 b appears to be metal-rich, even when compared to its slightly metal-rich host star \citep[M/H=0.10-0.28;][]{1993A&A...275..101E,2004A&A...418..551M,2005ApJS..159..141V,2007PASJ...59..335T,2010MNRAS.403.1368G,2012A&A...541A..40M,2013ApJ...764...78R}.  Based on Figure \ref{model parameters}, the high metallicity has a large effect on GJ 504 b's bright Ks photometry, which explains the object's unusual placement on the H-Ks vs. H color-magnitude diagram in Figure \ref{cmd}.  

The ability to measure the metallicity of individual planets, whether by broad-band photometry or by line-resolving spectroscopy, is hugely important for understanding the formation and evolution of extrasolar planets.  Core-accretion theory predicts that gas giant exoplanets should have higher metallicities than their host stars due to the infall of planetessimals \citep{1986Icar...67..409P,1988Icar...73..163P}.  In some circumstances, gravitational instability can also produce planets with metallicities that vary from their host stars \citep{2010ApJ...724..618B}, and indeed, higher metallicities may be favored \citep{2015MNRAS.448L..25N}.  While planet formation can create an object whose metallicity differs from its host star, binary star formation will not generally create objects with vastly different metallicities \citep{2004AA...420..683D,2006AA...454..581D}.  Therefore, within the confines of our atmospheric modeling, it appears that GJ 504 b formed like a planet, not like a binary star.

\subsubsection{Clouds}
GJ 504 b appears to have some cloud opacity.  Although dense silicate clouds are not seen in objects as cold as GJ 504 b, other condensates are predicted to affect the SEDs of cool atmospheres \citep{2012ApJ...756..172M}.  Low surface-gravity objects can support clouds at lower atmospheric pressures than high surface-gravity objects \citep{2012ApJ...754..135M}, although it remains to be seen if GJ 504 b's cloud properties are inconsistent with the cloud properties of more massive field brown dwarfs.

\subsubsection{Surface Gravity and Age}
Our analysis shows a preference for low surface gravities, constrained at the low end by evolutionary models, rather than atmosphere models (see our Bayesian prior in Section 4.2).  A low surface gravity suggests that GJ 504 b is a young, low-mass planet \citep{2013ApJ...774...11K} rather than an older and more massive brown dwarf \citep{2015ApJ...806..163F}, however, our posterior distribution does not completely rule out the later.  

If the age from \citet{2015ApJ...806..163F} is adopted, the correlations in Figure \ref{contour} show that the planet's cloud thickness is decreased, and its metallicity is increased.  To quantify this, we recalculate the posterior probabilities with the Bayesian prior that g$>$475$m/s^{2}$ .  We find T$_{eff}$=533$\pm$8 K, f$_{sed}$ (cloudiness)=4-5, [M/H]=0.78$\pm$0.08, R=0.95$\pm$0.06 R$_{Jup}$, and log(L)=-6.18$\pm$0.02 L$_{\Sun}$.

\subsubsection{Methane Absorption}
All of our models predict strong methane absorption at 1.66 and 3.3$\micron$, which is well-matched by the photometry.  Some warmer extrasolar planets (e.g., HR 8799 bcde and 2M1207 b) have effective temperatures where equilibrium chemistry models predict methane absorption \citep{2008Sci...322.1348M,2011ApJ...733...65B,2011ApJ...735L..39B}.  However, these planets show limited signs of methane absorption in near-infrared spectra \citep{2010AA...517A..76P,2011ApJ...733...65B,2013Sci...339.1398K,2015ApJ...804...61B}, or in mid-infrared SEDs \citep{2010ApJ...716..417H,2012ApJ...753...14S,2014ApJ...792...17S}, indicating the presence of non-equilibrium chemistry in the CH$_{4}\leftrightarrow$CO reaction network.  For cooler planets, like GJ 504 b, non-equilibrium chemistry is not expected to suppress methane absorption \citep{2014ApJ...797...41Z}.  The photometric upper limit to the brightness of GJ 504 b at 1.66~\micron \citep[$> 20.62 (3\sigma)$; ][]{2013ApJ...778L...4J} indicates the presence of methane opacity.  Our narrow L-band photometry confirms that methane opacity is affecting the slope of the 3-4$\micron$ SED in a way that is consistent with equilibrium chemistry models.

\section{Summary and Conclusions}
We obtained images of what is currently the coldest directly imaged exoplanet, GJ 504 b, in 3 narrow L-band filters, mapping out the 3.3$\micron$ methane fundamental absorption feature, and putting further constraints on the planet's basic physical properties.  With a best-fit temperature of 550 K, GJ 504 b is analogous to field brown dwarfs with a T spectral type.  Indeed, the SED of GJ 504 b shows many similarities to a T-type brown dwarf: strong methane absorption features at 1.66 and 3.3$\micron$, blue J-H colors that imply a relative lack of clouds compared to red L-dwarfs, and an SED that is reasonably well fit by a water-vapor dominated spectrum.  On the other hand, GJ 504 b's SED is different than any currently known field brown dwarf, particularly with regards to its unusually red H-Ks colors.  

We constructed a model grid with radius, temperature, metallicity, surface gravity, and cloud types as free parameters to try to explain GJ 504 b's unusual SED.  We find T$_{eff}$=544$\pm$10 K, f$_{sed}$ (cloudiness)=2-5, g$<$600 $m/s^{2}$, [M/H]=0.6$\pm$0.12, R=0.96$\pm$0.07 R$_{Jup}$, log(L)=-6.13$\pm$0.03 L$_{\Sun}$.  If GJ 504 is young (0.1-0.5 Gyr), \citet{2003A&A...402..701B} models predict that the companion is 3-8 M$_{jup}$.  If GJ 504 is old (3-6.5 Gyr), \citet{2003A&A...402..701B} models predict that the companion is 19-30 M$_{jup}$.  Our estimate of the planet's surface gravity favors the low-mass interpretation, but not conclusively.

\begin{figure}
\begin{center}
\includegraphics[angle=90,width=\columnwidth]{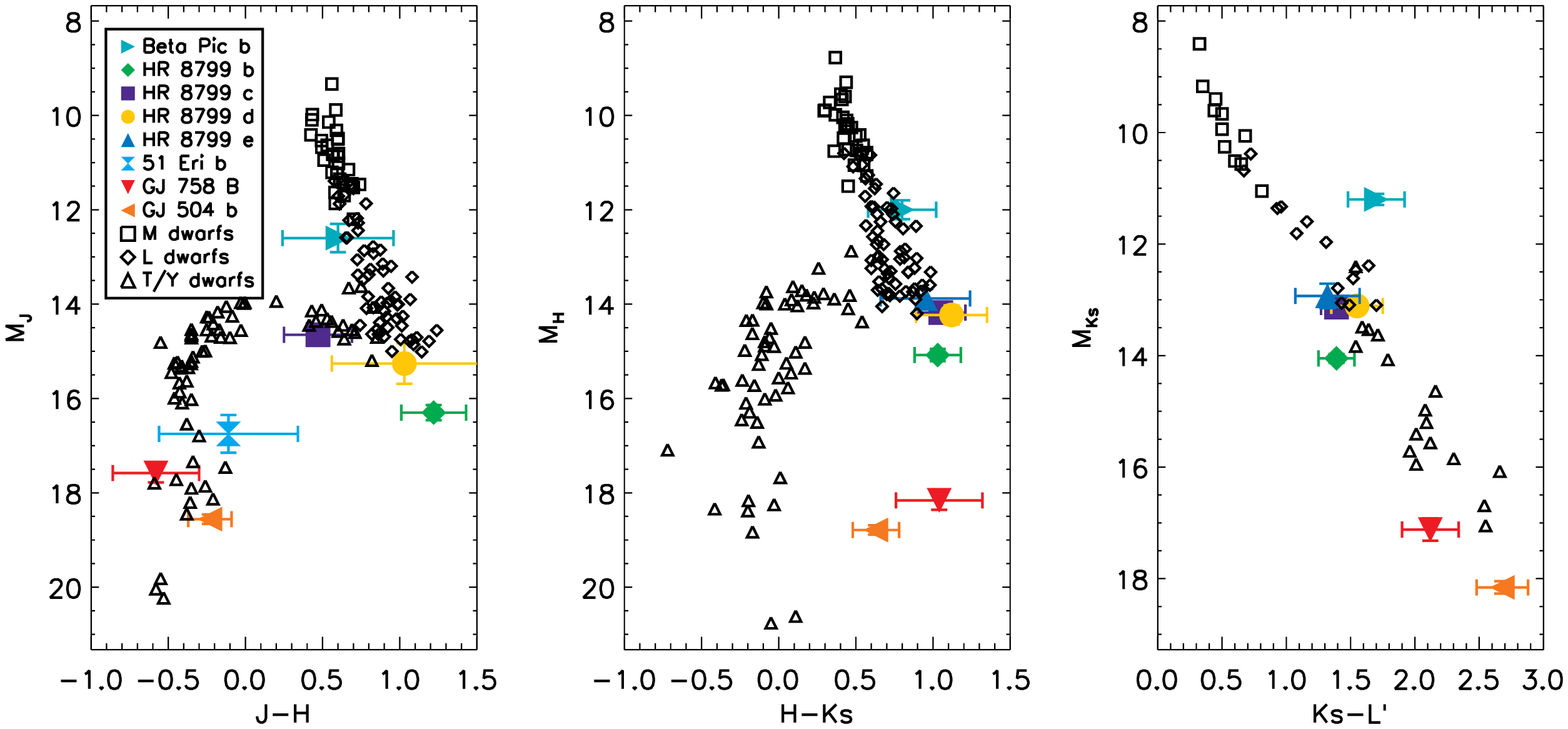}
\caption{Color-magnitude diagrams showing the field M$\rightarrow$L$\rightarrow$T brown dwarf sequence and directly-imaged exoplanets.  In the middle diagram (H-Ks vs. H), GJ 504 b is highly discrepant with similar magnitude field brown dwarfs.  The directly imaged brown dwarf, GJ 758 B has a similar appearance.  Photometry is compiled from \citet{2011AA...528L..15B,2013AA...555A.107B,2008Sci...322.1348M,2010Natur.468.1080M,2012ApJ...753...14S,2013ApJ...774...11K,2013ApJ...778L...4J,2011ApJ...728...85J,2012ApJS..201...19D} and \citet{2015Sci...350...64M} with some adjustments described in Section \ref{photometry section} (note that GJ 758 B's K photometry is Kc, not Ks). The brown dwarf photometry has been selected to have errors smaller than 0.1 mags in each filter, and uses K$_{MKO}$ instead of Ks, which is more complete for later spectral types and produces qualitatively similar results.
\label{cmd}}
\end{center}
\end{figure}

Of particular note, our planet atmosphere model requires a super-stellar metallicity to explain GJ 504 b's complete SED, and particularly its H-Ks colors shown in Figure \ref{cmd}.  Various planet formation models predict that planet metallicities should differ from their host stars.  Conversely, pairs of objects that form by binary fragmentation should have similar metallicities.  Within the confines of our atmospheric models, this suggests that GJ 504 b formed like a planet, not like a binary companion.  This result is independent of GJ 504 b's age and mass.

Additional photometry and spectroscopy at higher signal-to-noise will further improve our ability to understand planets like GJ 504 b.  In particular, observations over a broad wavelength range, including the mid-infrared ($\gtrsim$3$\micron$), can help break degeneracies in model parameters.  Spectroscopy in the mid-infrared will be particularly valuable \citep{2015arXiv150806290S}.  The \textit{James Webb Space Telescope} will discover and characterize a variety of new worlds \citep[e.g.,][]{2010PASP..122..162B}.  The LEECH exoplanet imaging survey is searching for and characterizing cool exoplanets in the mid-infrared right now, with the goal of improving our theoretical understanding of their atmospheres so that we can take full advantage of JWST's limited lifespan.

\acknowledgements
The authors thank the anonymous referee for a very helpful report.  We thank the HiCIAO team and Masayuki Kuzuhara for providing filter curves for the HiCIAO methane filters.  A.S. is supported by the National Aeronautics and Space Administration through Hubble Fellowship grant HSTHF2-51349 awarded by the Space Telescope Science Institute, which is operated by the Association of Universities for Research in Astronomy, Inc., for NASA, under contract NAS 5-26555.  E.B. is supported by the Swiss National Science Foundation (SNSF).  S.E., S.D., and A.L.M. acknowledge support from the ``Progetti Premiali'' funding scheme of the Italian Ministry of Education, University, and Research.  C.V. is supported by NSF AST-1312305.  LEECH is funded by the NASA Origins of Solar Systems Program, grant NNX13AJ17G.  This material is based in part upon work supported by the National Aeronautics and Space Administration under Agreement No. NNX15AD94G, Earths in Other Solar Systems, issued through the Science Mission Directorate interdivisional initiative Nexus for Exoplanet System Science.  The Large Binocular Telescope Interferometer is funded by the National Aeronautics and Space Administration as part of its Exoplanet Exploration program.  LMIRcam is funded by the National Science Foundation through grant NSF AST-0705296.  This publication makes use of the \textit{Spex Prism Library Analysis Toolkit}.  This publication makes use of data products from the Wide-field Infrared Survey Explorer, which is a joint project of the University of California, Los Angeles, and the Jet Propulsion Laboratory/California Institute of Technology, and NEOWISE, which is a project of the Jet Propulsion Laboratory/California Institute of Technology. WISE and NEOWISE are funded by the National Aeronautics and Space Administration.

\clearpage

\clearpage

\clearpage

\clearpage

\clearpage

\clearpage

\clearpage

\clearpage

\begin{deluxetable}{lcccccccccccc}
\tabletypesize{\scriptsize}
\tablecaption{Filter Properties}
\tablewidth{0pt}
\tablehead{
\colhead{} &
\colhead{$\lambda_{eff}$ ($\micron$)} &
\colhead{FWHM ($\micron$)} &
\colhead{zero-point flux (Jy)}
}
\startdata
L$_{NB6}$                  & 3.71 & 0.19 & 257 \\
L$_{NB7}$                  & 3.88 & 0.23 & 239 \\
L$_{NB8}$                  & 4.00 & 0.06 & 224 \\
\enddata
\tablecomments{Manufacturer's curves for $L_{NB1}$-$L_{NB7}$ were previously used and tabulated in \citet{2014ApJ...792...17S}.  L$_{NB8}$ is an additional filter at longer wavelength.  Here, we include the filters' cryogenic shifts based on an on-sky wavelength calibration of the LMIRcam grism \citep{2014DPS....4641815S}}
\label{filter table}
\end{deluxetable}

\clearpage

\begin{deluxetable}{lcccccccccccc}
\tabletypesize{\scriptsize}
\tablecaption{Observations}
\tablewidth{0pt}
\tablehead{
\colhead{Date} &
\colhead{Filter} &
\colhead{Aperture} &
\colhead{Frame Time (sat/unsat)} &
\colhead{Int Time} &
\colhead{Conditions}
\\
\colhead{} &
\colhead{} &
\colhead{} &
\colhead{(seconds)} &
\colhead{(minutes)} &
\colhead{} &

}
\startdata
UT 2013 April 21 & L$_{NB7}$ & 8.4m          & 0.524/0.058 & 55  & photometric, 0.9'' seeing \\
UT 2014 March 11 & L$_{NB8}$ & 2$\times$8.4m & 0.990/0.087 & 30  & patchy clouds, 1.4'' seeing \\
UT 2014 March 12 & L$_{NB6}$ & 2$\times$8.4m & 0.291/0.029 & 101 & photometric, 0.9'' seeing \\
UT 2014 March 13 & L$_{NB8}$ & 2$\times$8.4m & 0.873/0.087 & 44  & patchy clouds and then clear, 1.0'' seeing \\
\enddata
\label{observations}
\end{deluxetable}

\clearpage

\begin{deluxetable}{lcccccccccccc}
\rotate
\tabletypesize{\scriptsize}
\tablecaption{GJ 504 Photometry (Apparent Magnitudes)}
\tablewidth{0pt}
\tablehead{
\colhead{} &
\colhead{J} &
\colhead{H} &
\colhead{$\rm{CH}_{4}\rm{s}$} &
\colhead{$\rm{CH}_{4}\rm{l}$} &
\colhead{Ks} &
\colhead{L'} &
\colhead{L$_{NB6}$} &               
\colhead{L$_{NB7}$} &               
\colhead{L$_{NB8}$} 
\\
\colhead{} &
\colhead{(1.25$\micron$)} &
\colhead{(1.63$\micron$)} &
\colhead{(1.56$\micron$)} &
\colhead{(1.71$\micron$)} &
\colhead{(2.15$\micron$)} &
\colhead{(3.78$\micron$)} &
\colhead{(3.71$\micron$)} &
\colhead{(3.88$\micron$)} &
\colhead{(4.00$\micron$)} &

}

\startdata
GJ 504 A         & 4.13$\pm$0.02  & 3.88$\pm$0.02  & 3.89$\pm$0.02  & 3.87$\pm$0.02        & 3.81$\pm$0.02  & 3.80$\pm$0.02  & 3.80$\pm$0.02  & 3.80$\pm$0.02  & 3.80$\pm$0.02  \\
GJ 504 b-A       & 15.65$\pm$0.10 & 16.13$\pm$0.10 & 15.71$\pm$0.12 & $>$16.77             & 15.57$\pm$0.11 & 12.90$\pm$0.17 & 13.79$\pm$0.17 & 12.67$\pm$0.19 & 12.05$\pm$0.17 \\
GJ 504 b         & 19.78$\pm$0.10 & 20.01$\pm$0.10 & 19.60$\pm$0.12 & $>$20.64 (3$\sigma$) & 19.38$\pm$0.11 & 16.70$\pm$0.17 & 17.59$\pm$0.17 & 16.47$\pm$0.19 & 15.85$\pm$0.17 \\
\enddata
\vspace{-0.8cm}
\tablecomments{The GJ 504 A photometry has been recalculated, as described in Section \ref{photometry section}.  Relative photometry is from \citet{2013ApJ...778L...4J} for J, H, CH4s, CH4l, and Ks, \citet{2013ApJ...774...11K} for L', and this work for L$_{NB6}$, L$_{NB7}$, and L$_{NB8}$.}
\label{photometry}
\end{deluxetable}

\clearpage

\bibliographystyle{apj}
\bibliography{database}

\begin{thebibliography}{83}
\expandafter\ifx\csname natexlab\endcsname\relax\def\natexlab#1{#1}\fi

\bibitem[{{Ackerman} \& {Marley}(2001)}]{2001ApJ...556..872A}
{Ackerman}, A.~S. \& {Marley}, M.~S. 2001, \apj, 556, 872

\bibitem[{{Allard} {et~al.}(2005){Allard}, {Allard}, \&
  {Kielkopf}}]{2005A&A...440.1195A}
{Allard}, N.~F., {Allard}, F., \& {Kielkopf}, J.~F. 2005, \aap, 440, 1195

\bibitem[{{Amara} \& {Quanz}(2012)}]{2012MNRAS.427..948A}
{Amara}, A. \& {Quanz}, S.~P. 2012, \mnras, 427, 948

\bibitem[{{Bailey} {et~al.}(2014){Bailey}, {Hinz}, {Puglisi}, {Esposito},
  {Vaitheeswaran}, {Skemer}, {Defr{\`e}re}, {Vaz}, \&
  {Leisenring}}]{2014SPIE.9148E..03B}
{Bailey}, V.~P., {Hinz}, P.~M., {Puglisi}, A.~T., {Esposito}, S.,
  {Vaitheeswaran}, V., {Skemer}, A.~J., {Defr{\`e}re}, D., {Vaz}, A., \&
  {Leisenring}, J.~M. 2014, in Society of Photo-Optical Instrumentation
  Engineers (SPIE) Conference Series, Vol. 9148, Society of Photo-Optical
  Instrumentation Engineers (SPIE) Conference Series, 3

\bibitem[{{Baraffe} {et~al.}(2003){Baraffe}, {Chabrier}, {Barman}, {Allard}, \&
  {Hauschildt}}]{2003A&A...402..701B}
{Baraffe}, I., {Chabrier}, G., {Barman}, T.~S., {Allard}, F., \& {Hauschildt},
  P.~H. 2003, \aap, 402, 701

\bibitem[{{Barman} {et~al.}(2015){Barman}, {Konopacky}, {Macintosh}, \&
  {Marois}}]{2015ApJ...804...61B}
{Barman}, T.~S., {Konopacky}, Q.~M., {Macintosh}, B., \& {Marois}, C. 2015,
  \apj, 804, 61

\bibitem[{{Barman} {et~al.}(2011{\natexlab{a}}){Barman}, {Macintosh},
  {Konopacky}, \& {Marois}}]{2011ApJ...733...65B}
{Barman}, T.~S., {Macintosh}, B., {Konopacky}, Q.~M., \& {Marois}, C.
  2011{\natexlab{a}}, \apj, 733, 65

\bibitem[{{Barman} {et~al.}(2011{\natexlab{b}}){Barman}, {Macintosh},
  {Konopacky}, \& {Marois}}]{2011ApJ...735L..39B}
---. 2011{\natexlab{b}}, \apjl, 735, L39

\bibitem[{{Beichman} {et~al.}(2010){Beichman}, {Krist}, {Trauger}, {Greene},
  {Oppenheimer}, {Sivaramakrishnan}, {Doyon}, {Boccaletti}, {Barman}, \&
  {Rieke}}]{2010PASP..122..162B}
{Beichman}, C.~A., {Krist}, J., {Trauger}, J.~T., {Greene}, T., {Oppenheimer},
  B., {Sivaramakrishnan}, A., {Doyon}, R., {Boccaletti}, A., {Barman}, T.~S.,
  \& {Rieke}, M. 2010, \pasp, 122, 162

\bibitem[{{Boley} \& {Durisen}(2010)}]{2010ApJ...724..618B}
{Boley}, A.~C. \& {Durisen}, R.~H. 2010, \apj, 724, 618

\bibitem[{{Bonnefoy} {et~al.}(2013){Bonnefoy}, {Boccaletti}, {Lagrange},
  {Allard}, {Mordasini}, {Beust}, {Chauvin}, {Girard}, {Homeier}, {Apai},
  {Lacour}, \& {Rouan}}]{2013AA...555A.107B}
{Bonnefoy}, M., {Boccaletti}, A., {Lagrange}, A.-M., {Allard}, F., {Mordasini},
  C., {Beust}, H., {Chauvin}, G., {Girard}, J.~H.~V., {Homeier}, D., {Apai},
  D., {Lacour}, S., \& {Rouan}, D. 2013, \aap, 555, A107

\bibitem[{{Bonnefoy} {et~al.}(2014){Bonnefoy}, {Currie}, {Marleau},
  {Schlieder}, {Wisniewski}, {Carson}, {Covey}, {Henning}, {Biller}, {Hinz},
  {Klahr}, {Marsh Boyer}, {Zimmerman}, {Janson}, {McElwain}, {Mordasini},
  {Skemer}, {Bailey}, {Defr{\`e}re}, {Thalmann}, {Skrutskie}, {Allard},
  {Homeier}, {Tamura}, {Feldt}, {Cumming}, {Grady}, {Brandner}, {Helling},
  {Witte}, {Hauschildt}, {Kandori}, {Kuzuhara}, {Fukagawa}, {Kwon}, {Kudo},
  {Hashimoto}, {Kusakabe}, {Abe}, {Brandt}, {Egner}, {Guyon}, {Hayano},
  {Hayashi}, {Hayashi}, {Hodapp}, {Ishii}, {Iye}, {Knapp}, {Matsuo}, {Mede},
  {Miyama}, {Morino}, {Moro-Martin}, {Nishimura}, {Pyo}, {Serabyn}, {Suenaga},
  {Suto}, {Suzuki}, {Takahashi}, {Takami}, {Takato}, {Terada}, {Tomono},
  {Turner}, {Watanabe}, {Yamada}, {Takami}, \& {Usuda}}]{2014AA...562A.111B}
{Bonnefoy}, M., {Currie}, T., {Marleau}, G.-D., {Schlieder}, J.~E.,
  {Wisniewski}, J., {Carson}, J., {Covey}, K.~R., {Henning}, T., {Biller}, B.,
  {Hinz}, P., {Klahr}, H., {Marsh Boyer}, A.~N., {Zimmerman}, N., {Janson}, M.,
  {McElwain}, M., {Mordasini}, C., {Skemer}, A., {Bailey}, V., {Defr{\`e}re},
  D., {Thalmann}, C., {Skrutskie}, M., {Allard}, F., {Homeier}, D., {Tamura},
  M., {Feldt}, M., {Cumming}, A., {Grady}, C., {Brandner}, W., {Helling}, C.,
  {Witte}, S., {Hauschildt}, P., {Kandori}, R., {Kuzuhara}, M., {Fukagawa}, M.,
  {Kwon}, J., {Kudo}, T., {Hashimoto}, J., {Kusakabe}, N., {Abe}, L., {Brandt},
  T., {Egner}, S., {Guyon}, O., {Hayano}, Y., {Hayashi}, M., {Hayashi}, S.,
  {Hodapp}, K., {Ishii}, M., {Iye}, M., {Knapp}, G., {Matsuo}, T., {Mede}, K.,
  {Miyama}, M., {Morino}, J.-I., {Moro-Martin}, A., {Nishimura}, T., {Pyo}, T.,
  {Serabyn}, E., {Suenaga}, T., {Suto}, H., {Suzuki}, R., {Takahashi},
  {Takami}, M., {Takato}, N., {Terada}, H., {Tomono}, D., {Turner}, E.,
  {Watanabe}, M., {Yamada}, T., {Takami}, H., \& {Usuda}, T. 2014, \aap, 562,
  A111

\bibitem[{{Bonnefoy} {et~al.}(2011){Bonnefoy}, {Lagrange}, {Boccaletti},
  {Chauvin}, {Apai}, {Allard}, {Ehrenreich}, {Girard}, {Mouillet}, {Rouan},
  {Gratadour}, \& {Kasper}}]{2011AA...528L..15B}
{Bonnefoy}, M., {Lagrange}, A.-M., {Boccaletti}, A., {Chauvin}, G., {Apai}, D.,
  {Allard}, F., {Ehrenreich}, D., {Girard}, J.~H.~V., {Mouillet}, D., {Rouan},
  D., {Gratadour}, D., \& {Kasper}, M. 2011, \aap, 528, L15

\bibitem[{{Carlberg} {et~al.}(2012){Carlberg}, {Cunha}, {Smith}, \&
  {Majewski}}]{2012ApJ...757..109C}
{Carlberg}, J.~K., {Cunha}, K., {Smith}, V.~V., \& {Majewski}, S.~R. 2012,
  \apj, 757, 109

\bibitem[{{Castelli} \& {Kurucz}(2004)}]{2004astro.ph..5087C}
{Castelli}, F. \& {Kurucz}, R.~L. 2004, ArXiv Astrophysics e-prints

\bibitem[{{Chauvin} {et~al.}(2004){Chauvin}, {Lagrange}, {Dumas}, {Zuckerman},
  {Mouillet}, {Song}, {Beuzit}, \& {Lowrance}}]{2004AA...425L..29C}
{Chauvin}, G., {Lagrange}, A.-M., {Dumas}, C., {Zuckerman}, B., {Mouillet}, D.,
  {Song}, I., {Beuzit}, J.-L., \& {Lowrance}, P. 2004, \aap, 425, L29

\bibitem[{{Cutri}(2013)}]{2013yCat.2328....0C}
{Cutri}, R.~M. e.~a. 2013, VizieR Online Data Catalog, 2328, 0

\bibitem[{{Desidera} {et~al.}(2006){Desidera}, {Gratton}, {Lucatello}, \&
  {Claudi}}]{2006AA...454..581D}
{Desidera}, S., {Gratton}, R.~G., {Lucatello}, S., \& {Claudi}, R.~U. 2006,
  \aap, 454, 581

\bibitem[{{Desidera} {et~al.}(2004){Desidera}, {Gratton}, {Scuderi}, {Claudi},
  {Cosentino}, {Barbieri}, {Bonanno}, {Carretta}, {Endl}, {Lucatello},
  {Martinez Fiorenzano}, \& {Marzari}}]{2004AA...420..683D}
{Desidera}, S., {Gratton}, R.~G., {Scuderi}, S., {Claudi}, R.~U., {Cosentino},
  R., {Barbieri}, M., {Bonanno}, G., {Carretta}, E., {Endl}, M., {Lucatello},
  S., {Martinez Fiorenzano}, A.~F., \& {Marzari}, F. 2004, \aap, 420, 683

\bibitem[{{Dupuy} \& {Liu}(2012)}]{2012ApJS..201...19D}
{Dupuy}, T.~J. \& {Liu}, M.~C. 2012, \apjs, 201, 19

\bibitem[{{Edvardsson} {et~al.}(1993){Edvardsson}, {Andersen}, {Gustafsson},
  {Lambert}, {Nissen}, \& {Tomkin}}]{1993A&A...275..101E}
{Edvardsson}, B., {Andersen}, J., {Gustafsson}, B., {Lambert}, D.~L., {Nissen},
  P.~E., \& {Tomkin}, J. 1993, \aap, 275, 101

\bibitem[{{Esposito} {et~al.}(2011){Esposito}, {Riccardi}, {Pinna}, {Puglisi},
  {Quir{\'o}s-Pacheco}, {Arcidiacono}, {Xompero}, {Briguglio}, {Agapito},
  {Busoni}, {Fini}, {Argomedo}, {Gherardi}, {Brusa}, {Miller}, {Guerra},
  {Stefanini}, \& {Salinari}}]{2011SPIE.8149E...1E}
{Esposito}, S., {Riccardi}, A., {Pinna}, E., {Puglisi}, A.,
  {Quir{\'o}s-Pacheco}, F., {Arcidiacono}, C., {Xompero}, M., {Briguglio}, R.,
  {Agapito}, G., {Busoni}, L., {Fini}, L., {Argomedo}, J., {Gherardi}, A.,
  {Brusa}, G., {Miller}, D., {Guerra}, J.~C., {Stefanini}, P., \& {Salinari},
  P. 2011, in Society of Photo-Optical Instrumentation Engineers (SPIE)
  Conference Series, Vol. 8149, Society of Photo-Optical Instrumentation
  Engineers (SPIE) Conference Series

\bibitem[{{Faherty} {et~al.}(2013){Faherty}, {Rice}, {Cruz}, {Mamajek}, \&
  {N{\'u}{\~n}ez}}]{2013AJ....145....2F}
{Faherty}, J.~K., {Rice}, E.~L., {Cruz}, K.~L., {Mamajek}, E.~E., \&
  {N{\'u}{\~n}ez}, A. 2013, \aj, 145, 2

\bibitem[{{Fergus} {et~al.}(2014){Fergus}, {Hogg}, {Oppenheimer}, {Brenner}, \&
  {Pueyo}}]{2014ApJ...794..161F}
{Fergus}, R., {Hogg}, D.~W., {Oppenheimer}, R., {Brenner}, D., \& {Pueyo}, L.
  2014, \apj, 794, 161

\bibitem[{{Fischer} \& {Valenti}(2005)}]{2005ApJ...622.1102F}
{Fischer}, D.~A. \& {Valenti}, J. 2005, \apj, 622, 1102

\bibitem[{{Fortney} {et~al.}(2007){Fortney}, {Marley}, \&
  {Barnes}}]{2007ApJ...659.1661F}
{Fortney}, J.~J., {Marley}, M.~S., \& {Barnes}, J.~W. 2007, \apj, 659, 1661

\bibitem[{{Fortney} {et~al.}(2008){Fortney}, {Marley}, {Saumon}, \&
  {Lodders}}]{2008ApJ...683.1104F}
{Fortney}, J.~J., {Marley}, M.~S., {Saumon}, D., \& {Lodders}, K. 2008, \apj,
  683, 1104

\bibitem[{{Fuhrmann} \& {Chini}(2015)}]{2015ApJ...806..163F}
{Fuhrmann}, K. \& {Chini}, R. 2015, \apj, 806, 163

\bibitem[{{Gauza} {et~al.}(2015){Gauza}, {B{\'e}jar}, {P{\'e}rez-Garrido},
  {Rosa Zapatero Osorio}, {Lodieu}, {Rebolo}, {Pall{\'e}}, \&
  {Nowak}}]{2015ApJ...804...96G}
{Gauza}, B., {B{\'e}jar}, V.~J.~S., {P{\'e}rez-Garrido}, A., {Rosa Zapatero
  Osorio}, M., {Lodieu}, N., {Rebolo}, R., {Pall{\'e}}, E., \& {Nowak}, G.
  2015, \apj, 804, 96

\bibitem[{{Gonzalez} {et~al.}(2010){Gonzalez}, {Carlson}, \&
  {Tobin}}]{2010MNRAS.403.1368G}
{Gonzalez}, G., {Carlson}, M.~K., \& {Tobin}, R.~W. 2010, \mnras, 403, 1368

\bibitem[{{Hinz} {et~al.}(2012){Hinz}, {Arbo}, {Bailey}, {Connors}, {Durney},
  {Esposito}, {Hoffmann}, {Jones}, {Leisenring}, {Montoya}, {Nash}, {Nelson},
  {McMahon}, {Pinna}, {Puglisi}, {Skemer}, {Skrutskie}, \&
  {Vaitheeswaran}}]{2012SPIE.8445E..0UH}
{Hinz}, P., {Arbo}, P., {Bailey}, V., {Connors}, T., {Durney}, O., {Esposito},
  S., {Hoffmann}, W., {Jones}, T., {Leisenring}, J., {Montoya}, M., {Nash}, M.,
  {Nelson}, M., {McMahon}, T., {Pinna}, E., {Puglisi}, A., {Skemer}, A.,
  {Skrutskie}, M., \& {Vaitheeswaran}, V. 2012, in Society of Photo-Optical
  Instrumentation Engineers (SPIE) Conference Series, Vol. 8445, Society of
  Photo-Optical Instrumentation Engineers (SPIE) Conference Series

\bibitem[{{Hinz} {et~al.}(2010){Hinz}, {Rodigas}, {Kenworthy}, {Sivanandam},
  {Heinze}, {Mamajek}, \& {Meyer}}]{2010ApJ...716..417H}
{Hinz}, P.~M., {Rodigas}, T.~J., {Kenworthy}, M.~A., {Sivanandam}, S.,
  {Heinze}, A.~N., {Mamajek}, E.~E., \& {Meyer}, M.~R. 2010, \apj, 716, 417

\bibitem[{{Janson} {et~al.}(2013){Janson}, {Brandt}, {Kuzuhara}, {Spiegel},
  {Thalmann}, {Currie}, {Bonnefoy}, {Zimmerman}, {Sorahana}, {Kotani},
  {Schlieder}, {Hashimoto}, {Kudo}, {Kusakabe}, {Abe}, {Brandner}, {Carson},
  {Egner}, {Feldt}, {Goto}, {Grady}, {Guyon}, {Hayano}, {Hayashi}, {Hayashi},
  {Henning}, {Hodapp}, {Ishii}, {Iye}, {Kandori}, {Knapp}, {Kwon}, {Matsuo},
  {McElwain}, {Mede}, {Miyama}, {Morino}, {Moro-Mart{\'{\i}}n}, {Nakagawa},
  {Nishimura}, {Pyo}, {Serabyn}, {Suenaga}, {Suto}, {Suzuki}, {Takahashi},
  {Takami}, {Takato}, {Terada}, {Tomono}, {Turner}, {Watanabe}, {Wisniewski},
  {Yamada}, {Takami}, {Usuda}, \& {Tamura}}]{2013ApJ...778L...4J}
{Janson}, M., {Brandt}, T.~D., {Kuzuhara}, M., {Spiegel}, D.~S., {Thalmann},
  C., {Currie}, T., {Bonnefoy}, M., {Zimmerman}, N., {Sorahana}, S., {Kotani},
  T., {Schlieder}, J., {Hashimoto}, J., {Kudo}, T., {Kusakabe}, N., {Abe}, L.,
  {Brandner}, W., {Carson}, J.~C., {Egner}, S., {Feldt}, M., {Goto}, M.,
  {Grady}, C.~A., {Guyon}, O., {Hayano}, Y., {Hayashi}, M., {Hayashi}, S.,
  {Henning}, T., {Hodapp}, K.~W., {Ishii}, M., {Iye}, M., {Kandori}, R.,
  {Knapp}, G.~R., {Kwon}, J., {Matsuo}, T., {McElwain}, M.~W., {Mede}, K.,
  {Miyama}, S., {Morino}, J.-I., {Moro-Mart{\'{\i}}n}, A., {Nakagawa}, T.,
  {Nishimura}, T., {Pyo}, T.-S., {Serabyn}, E., {Suenaga}, T., {Suto}, H.,
  {Suzuki}, R., {Takahashi}, Y., {Takami}, M., {Takato}, N., {Terada}, H.,
  {Tomono}, D., {Turner}, E.~L., {Watanabe}, M., {Wisniewski}, J., {Yamada},
  T., {Takami}, H., {Usuda}, T., \& {Tamura}, M. 2013, \apjl, 778, L4

\bibitem[{{Janson} {et~al.}(2011){Janson}, {Carson}, {Thalmann}, {McElwain},
  {Goto}, {Crepp}, {Wisniewski}, {Abe}, {Brandner}, {Burrows}, {Egner},
  {Feldt}, {Grady}, {Golota}, {Guyon}, {Hashimoto}, {Hayano}, {Hayashi},
  {Hayashi}, {Henning}, {Hodapp}, {Ishii}, {Iye}, {Kandori}, {Knapp}, {Kudo},
  {Kusakabe}, {Kuzuhara}, {Matsuo}, {Mayama}, {Miyama}, {Morino},
  {Moro-Mart{\'{\i}}n}, {Nishimura}, {Pyo}, {Serabyn}, {Suto}, {Suzuki},
  {Takami}, {Takato}, {Terada}, {Tofflemire}, {Tomono}, {Turner}, {Watanabe},
  {Yamada}, {Takami}, {Usuda}, \& {Tamura}}]{2011ApJ...728...85J}
{Janson}, M., {Carson}, J., {Thalmann}, C., {McElwain}, M.~W., {Goto}, M.,
  {Crepp}, J., {Wisniewski}, J., {Abe}, L., {Brandner}, W., {Burrows}, A.,
  {Egner}, S., {Feldt}, M., {Grady}, C.~A., {Golota}, T., {Guyon}, O.,
  {Hashimoto}, J., {Hayano}, Y., {Hayashi}, M., {Hayashi}, S., {Henning}, T.,
  {Hodapp}, K.~W., {Ishii}, M., {Iye}, M., {Kandori}, R., {Knapp}, G.~R.,
  {Kudo}, T., {Kusakabe}, N., {Kuzuhara}, M., {Matsuo}, T., {Mayama}, S.,
  {Miyama}, S., {Morino}, J.-I., {Moro-Mart{\'{\i}}n}, A., {Nishimura}, T.,
  {Pyo}, T.-S., {Serabyn}, E., {Suto}, H., {Suzuki}, R., {Takami}, M.,
  {Takato}, N., {Terada}, H., {Tofflemire}, B., {Tomono}, D., {Turner}, E.~L.,
  {Watanabe}, M., {Yamada}, T., {Takami}, H., {Usuda}, T., \& {Tamura}, M.
  2011, \apj, 728, 85

\bibitem[{{Kidger} \& {Mart{\'{\i}}n-Luis}(2003)}]{2003AJ....125.3311K}
{Kidger}, M.~R. \& {Mart{\'{\i}}n-Luis}, F. 2003, \aj, 125, 3311

\bibitem[{{Konopacky} {et~al.}(2013){Konopacky}, {Barman}, {Macintosh}, \&
  {Marois}}]{2013Sci...339.1398K}
{Konopacky}, Q.~M., {Barman}, T.~S., {Macintosh}, B.~A., \& {Marois}, C. 2013,
  Science, 339, 1398

\bibitem[{{Kuzuhara} {et~al.}(2013){Kuzuhara}, {Tamura}, {Kudo}, {Janson},
  {Kandori}, {Brandt}, {Thalmann}, {Spiegel}, {Biller}, {Carson}, {Hori},
  {Suzuki}, {Burrows}, {Henning}, {Turner}, {McElwain}, {Moro-Mart{\'{\i}}n},
  {Suenaga}, {Takahashi}, {Kwon}, {Lucas}, {Abe}, {Brandner}, {Egner}, {Feldt},
  {Fujiwara}, {Goto}, {Grady}, {Guyon}, {Hashimoto}, {Hayano}, {Hayashi},
  {Hayashi}, {Hodapp}, {Ishii}, {Iye}, {Knapp}, {Matsuo}, {Mayama}, {Miyama},
  {Morino}, {Nishikawa}, {Nishimura}, {Kotani}, {Kusakabe}, {Pyo}, {Serabyn},
  {Suto}, {Takami}, {Takato}, {Terada}, {Tomono}, {Watanabe}, {Wisniewski},
  {Yamada}, {Takami}, \& {Usuda}}]{2013ApJ...774...11K}
{Kuzuhara}, M., {Tamura}, M., {Kudo}, T., {Janson}, M., {Kandori}, R.,
  {Brandt}, T.~D., {Thalmann}, C., {Spiegel}, D., {Biller}, B., {Carson}, J.,
  {Hori}, Y., {Suzuki}, R., {Burrows}, A., {Henning}, T., {Turner}, E.~L.,
  {McElwain}, M.~W., {Moro-Mart{\'{\i}}n}, A., {Suenaga}, T., {Takahashi},
  Y.~H., {Kwon}, J., {Lucas}, P., {Abe}, L., {Brandner}, W., {Egner}, S.,
  {Feldt}, M., {Fujiwara}, H., {Goto}, M., {Grady}, C.~A., {Guyon}, O.,
  {Hashimoto}, J., {Hayano}, Y., {Hayashi}, M., {Hayashi}, S.~S., {Hodapp},
  K.~W., {Ishii}, M., {Iye}, M., {Knapp}, G.~R., {Matsuo}, T., {Mayama}, S.,
  {Miyama}, S., {Morino}, J.-I., {Nishikawa}, J., {Nishimura}, T., {Kotani},
  T., {Kusakabe}, N., {Pyo}, T.-S., {Serabyn}, E., {Suto}, H., {Takami}, M.,
  {Takato}, N., {Terada}, H., {Tomono}, D., {Watanabe}, M., {Wisniewski},
  J.~P., {Yamada}, T., {Takami}, H., \& {Usuda}, T. 2013, \apj, 774, 11

\bibitem[{{Lafreni{\`e}re} {et~al.}(2007){Lafreni{\`e}re}, {Marois}, {Doyon},
  {Nadeau}, \& {Artigau}}]{2007ApJ...660..770L}
{Lafreni{\`e}re}, D., {Marois}, C., {Doyon}, R., {Nadeau}, D., \& {Artigau},
  {\'E}. 2007, \apj, 660, 770

\bibitem[{{Leisenring} {et~al.}(2012){Leisenring}, {Skrutskie}, {Hinz},
  {Skemer}, {Bailey}, {Eisner}, {Garnavich}, {Hoffmann}, {Jones}, {Kenworthy},
  {Kuzmenko}, {Meyer}, {Nelson}, {Rodigas}, {Wilson}, \&
  {Vaitheeswaran}}]{2012SPIE.8446E..4FL}
{Leisenring}, J.~M., {Skrutskie}, M.~F., {Hinz}, P.~M., {Skemer}, A., {Bailey},
  V., {Eisner}, J., {Garnavich}, P., {Hoffmann}, W.~F., {Jones}, T.,
  {Kenworthy}, M., {Kuzmenko}, P., {Meyer}, M., {Nelson}, M., {Rodigas}, T.~J.,
  {Wilson}, J.~C., \& {Vaitheeswaran}, V. 2012, in Society of Photo-Optical
  Instrumentation Engineers (SPIE) Conference Series, Vol. 8446, Society of
  Photo-Optical Instrumentation Engineers (SPIE) Conference Series

\bibitem[{{Liu} {et~al.}(2013){Liu}, {Magnier}, {Deacon}, {Allers}, {Dupuy},
  {Kotson}, {Aller}, {Burgett}, {Chambers}, {Draper}, {Hodapp}, {Jedicke},
  {Kaiser}, {Kudritzki}, {Metcalfe}, {Morgan}, {Price}, {Tonry}, \&
  {Wainscoat}}]{2013ApJ...777L..20L}
{Liu}, M.~C., {Magnier}, E.~A., {Deacon}, N.~R., {Allers}, K.~N., {Dupuy},
  T.~J., {Kotson}, M.~C., {Aller}, K.~M., {Burgett}, W.~S., {Chambers}, K.~C.,
  {Draper}, P.~W., {Hodapp}, K.~W., {Jedicke}, R., {Kaiser}, N., {Kudritzki},
  R.-P., {Metcalfe}, N., {Morgan}, J.~S., {Price}, P.~A., {Tonry}, J.~L., \&
  {Wainscoat}, R.~J. 2013, \apjl, 777, L20

\bibitem[{{Lloyd-Hart}(2000)}]{2000PASP..112..264L}
{Lloyd-Hart}, M. 2000, \pasp, 112, 264

\bibitem[{{Lodders}(1999)}]{1999ApJ...519..793L}
{Lodders}, K. 1999, \apj, 519, 793

\bibitem[{{Lodders}(2002)}]{2002ApJ...577..974L}
---. 2002, \apj, 577, 974

\bibitem[{{Lodders} \& {Fegley}(2002)}]{2002Icar..155..393L}
{Lodders}, K. \& {Fegley}, B. 2002, \icarus, 155, 393

\bibitem[{{Lodders} \& {Fegley}(2006)}]{2006asup.book....1L}
{Lodders}, K. \& {Fegley}, Jr., B. {Chemistry of Low Mass Substellar Objects},
  ed. J.~W. {Mason}, 1

\bibitem[{{Macintosh} {et~al.}(2015){Macintosh}, {Graham}, {Barman}, {De Rosa},
  {Konopacky}, {Marley}, {Marois}, {Nielsen}, {Pueyo}, {Rajan}, {Rameau},
  {Saumon}, {Wang}, {Patience}, {Ammons}, {Arriaga}, {Artigau}, {Beckwith},
  {Brewster}, {Bruzzone}, {Bulger}, {Burningham}, {Burrows}, {Chen}, {Chiang},
  {Chilcote}, {Dawson}, {Dong}, {Doyon}, {Draper}, {Duch{\^e}ne}, {Esposito},
  {Fabrycky}, {Fitzgerald}, {Follette}, {Fortney}, {Gerard}, {Goodsell},
  {Greenbaum}, {Hibon}, {Hinkley}, {Cotten}, {Hung}, {Ingraham},
  {Johnson-Groh}, {Kalas}, {Lafreniere}, {Larkin}, {Lee}, {Line}, {Long},
  {Maire}, {Marchis}, {Matthews}, {Max}, {Metchev}, {Millar-Blanchaer},
  {Mittal}, {Morley}, {Morzinski}, {Murray-Clay}, {Oppenheimer}, {Palmer},
  {Patel}, {Perrin}, {Poyneer}, {Rafikov}, {Rantakyr{\"o}}, {Rice}, {Rojo},
  {Rudy}, {Ruffio}, {Ruiz}, {Sadakuni}, {Saddlemyer}, {Salama}, {Savransky},
  {Schneider}, {Sivaramakrishnan}, {Song}, {Soummer}, {Thomas}, {Vasisht},
  {Wallace}, {Ward-Duong}, {Wiktorowicz}, {Wolff}, \&
  {Zuckerman}}]{2015Sci...350...64M}
{Macintosh}, B., {Graham}, J.~R., {Barman}, T., {De Rosa}, R.~J., {Konopacky},
  Q., {Marley}, M.~S., {Marois}, C., {Nielsen}, E.~L., {Pueyo}, L., {Rajan},
  A., {Rameau}, J., {Saumon}, D., {Wang}, J.~J., {Patience}, J., {Ammons}, M.,
  {Arriaga}, P., {Artigau}, E., {Beckwith}, S., {Brewster}, J., {Bruzzone}, S.,
  {Bulger}, J., {Burningham}, B., {Burrows}, A.~S., {Chen}, C., {Chiang}, E.,
  {Chilcote}, J.~K., {Dawson}, R.~I., {Dong}, R., {Doyon}, R., {Draper}, Z.~H.,
  {Duch{\^e}ne}, G., {Esposito}, T.~M., {Fabrycky}, D., {Fitzgerald}, M.~P.,
  {Follette}, K.~B., {Fortney}, J.~J., {Gerard}, B., {Goodsell}, S.,
  {Greenbaum}, A.~Z., {Hibon}, P., {Hinkley}, S., {Cotten}, T.~H., {Hung},
  L.-W., {Ingraham}, P., {Johnson-Groh}, M., {Kalas}, P., {Lafreniere}, D.,
  {Larkin}, J.~E., {Lee}, J., {Line}, M., {Long}, D., {Maire}, J., {Marchis},
  F., {Matthews}, B.~C., {Max}, C.~E., {Metchev}, S., {Millar-Blanchaer},
  M.~A., {Mittal}, T., {Morley}, C.~V., {Morzinski}, K.~M., {Murray-Clay}, R.,
  {Oppenheimer}, R., {Palmer}, D.~W., {Patel}, R., {Perrin}, M.~D., {Poyneer},
  L.~A., {Rafikov}, R.~R., {Rantakyr{\"o}}, F.~T., {Rice}, E.~L., {Rojo}, P.,
  {Rudy}, A.~R., {Ruffio}, J.-B., {Ruiz}, M.~T., {Sadakuni}, N., {Saddlemyer},
  L., {Salama}, M., {Savransky}, D., {Schneider}, A.~C., {Sivaramakrishnan},
  A., {Song}, I., {Soummer}, R., {Thomas}, S., {Vasisht}, G., {Wallace}, J.~K.,
  {Ward-Duong}, K., {Wiktorowicz}, S.~J., {Wolff}, S.~G., \& {Zuckerman}, B.
  2015, Science, 350, 64

\bibitem[{{Maire} {et~al.}(2015){Maire}, {Skemer}, {Hinz}, {Desidera},
  {Esposito}, {Gratton}, {Marzari}, {Skrutskie}, {Biller}, {Defr{\`e}re},
  {Bailey}, {Leisenring}, {Apai}, {Bonnefoy}, {Brandner}, {Buenzli}, {Claudi},
  {Close}, {Crepp}, {De Rosa}, {Eisner}, {Fortney}, {Henning}, {Hofmann},
  {Kopytova}, {Males}, {Mesa}, {Morzinski}, {Oza}, {Patience}, {Pinna},
  {Rajan}, {Schertl}, {Schlieder}, {Su}, {Vaz}, {Ward-Duong}, {Weigelt}, \&
  {Woodward}}]{2015A&A...576A.133M}
{Maire}, A.-L., {Skemer}, A.~J., {Hinz}, P.~M., {Desidera}, S., {Esposito}, S.,
  {Gratton}, R., {Marzari}, F., {Skrutskie}, M.~F., {Biller}, B.~A.,
  {Defr{\`e}re}, D., {Bailey}, V.~P., {Leisenring}, J.~M., {Apai}, D.,
  {Bonnefoy}, M., {Brandner}, W., {Buenzli}, E., {Claudi}, R.~U., {Close},
  L.~M., {Crepp}, J.~R., {De Rosa}, R.~J., {Eisner}, J.~A., {Fortney}, J.~J.,
  {Henning}, T., {Hofmann}, K.-H., {Kopytova}, T.~G., {Males}, J.~R., {Mesa},
  D., {Morzinski}, K.~M., {Oza}, A., {Patience}, J., {Pinna}, E., {Rajan}, A.,
  {Schertl}, D., {Schlieder}, J.~E., {Su}, K.~Y.~L., {Vaz}, A., {Ward-Duong},
  K., {Weigelt}, G., \& {Woodward}, C.~E. 2015, \aap, 576, A133

\bibitem[{{Maldonado} {et~al.}(2012){Maldonado}, {Eiroa}, {Villaver},
  {Montesinos}, \& {Mora}}]{2012A&A...541A..40M}
{Maldonado}, J., {Eiroa}, C., {Villaver}, E., {Montesinos}, B., \& {Mora}, A.
  2012, \aap, 541, A40

\bibitem[{{Mamajek} \& {Hillenbrand}(2008)}]{2008ApJ...687.1264M}
{Mamajek}, E.~E. \& {Hillenbrand}, L.~A. 2008, \apj, 687, 1264

\bibitem[{{Marley} {et~al.}(2012){Marley}, {Saumon}, {Cushing}, {Ackerman},
  {Fortney}, \& {Freedman}}]{2012ApJ...754..135M}
{Marley}, M.~S., {Saumon}, D., {Cushing}, M., {Ackerman}, A.~S., {Fortney},
  J.~J., \& {Freedman}, R. 2012, \apj, 754, 135

\bibitem[{{Marois} {et~al.}(2006){Marois}, {Lafreni{\`e}re}, {Doyon},
  {Macintosh}, \& {Nadeau}}]{2006ApJ...641..556M}
{Marois}, C., {Lafreni{\`e}re}, D., {Doyon}, R., {Macintosh}, B., \& {Nadeau},
  D. 2006, \apj, 641, 556

\bibitem[{{Marois} {et~al.}(2008){Marois}, {Macintosh}, {Barman}, {Zuckerman},
  {Song}, {Patience}, {Lafreni{\`e}re}, \& {Doyon}}]{2008Sci...322.1348M}
{Marois}, C., {Macintosh}, B., {Barman}, T., {Zuckerman}, B., {Song}, I.,
  {Patience}, J., {Lafreni{\`e}re}, D., \& {Doyon}, R. 2008, Science, 322, 1348

\bibitem[{{Marois} {et~al.}(2010){Marois}, {Zuckerman}, {Konopacky},
  {Macintosh}, \& {Barman}}]{2010Natur.468.1080M}
{Marois}, C., {Zuckerman}, B., {Konopacky}, Q.~M., {Macintosh}, B., \&
  {Barman}, T. 2010, \nat, 468, 1080

\bibitem[{{Mishenina} {et~al.}(2004){Mishenina}, {Soubiran}, {Kovtyukh}, \&
  {Korotin}}]{2004A&A...418..551M}
{Mishenina}, T.~V., {Soubiran}, C., {Kovtyukh}, V.~V., \& {Korotin}, S.~A.
  2004, \aap, 418, 551

\bibitem[{{Morley} {et~al.}(2012){Morley}, {Fortney}, {Marley}, {Visscher},
  {Saumon}, \& {Leggett}}]{2012ApJ...756..172M}
{Morley}, C.~V., {Fortney}, J.~J., {Marley}, M.~S., {Visscher}, C., {Saumon},
  D., \& {Leggett}, S.~K. 2012, \apj, 756, 172

\bibitem[{{Morley} {et~al.}(2014){Morley}, {Marley}, {Fortney}, {Lupu},
  {Saumon}, {Greene}, \& {Lodders}}]{2014ApJ...787...78M}
{Morley}, C.~V., {Marley}, M.~S., {Fortney}, J.~J., {Lupu}, R., {Saumon}, D.,
  {Greene}, T., \& {Lodders}, K. 2014, \apj, 787, 78

\bibitem[{{Moses} {et~al.}(2013){Moses}, {Madhusudhan}, {Visscher}, \&
  {Freedman}}]{2013ApJ...763...25M}
{Moses}, J.~I., {Madhusudhan}, N., {Visscher}, C., \& {Freedman}, R.~S. 2013,
  \apj, 763, 25

\bibitem[{{Nayakshin}(2015)}]{2015MNRAS.448L..25N}
{Nayakshin}, S. 2015, \mnras, 448, L25

\bibitem[{{Patience} {et~al.}(2010){Patience}, {King}, {de Rosa}, \&
  {Marois}}]{2010AA...517A..76P}
{Patience}, J., {King}, R.~R., {de Rosa}, R.~J., \& {Marois}, C. 2010, \aap,
  517, A76

\bibitem[{{Patience} {et~al.}(2012){Patience}, {King}, {De Rosa}, {Vigan},
  {Witte}, {Rice}, {Helling}, \& {Hauschildt}}]{2012A&A...540A..85P}
{Patience}, J., {King}, R.~R., {De Rosa}, R.~J., {Vigan}, A., {Witte}, S.,
  {Rice}, E., {Helling}, C., \& {Hauschildt}, P. 2012, \aap, 540, A85

\bibitem[{{Podolak} {et~al.}(1988){Podolak}, {Pollack}, \&
  {Reynolds}}]{1988Icar...73..163P}
{Podolak}, M., {Pollack}, J.~B., \& {Reynolds}, R.~T. 1988, \icarus, 73, 163

\bibitem[{{Pollack} {et~al.}(1986){Pollack}, {Podolak}, {Bodenheimer}, \&
  {Christofferson}}]{1986Icar...67..409P}
{Pollack}, J.~B., {Podolak}, M., {Bodenheimer}, P., \& {Christofferson}, B.
  1986, \icarus, 67, 409

\bibitem[{{Ram{\'{\i}}rez} {et~al.}(2013){Ram{\'{\i}}rez}, {Allende Prieto}, \&
  {Lambert}}]{2013ApJ...764...78R}
{Ram{\'{\i}}rez}, I., {Allende Prieto}, C., \& {Lambert}, D.~L. 2013, \apj,
  764, 78

\bibitem[{{Rieke} {et~al.}(2008){Rieke}, {Blaylock}, {Decin}, {Engelbracht},
  {Ogle}, {Avrett}, {Carpenter}, {Cutri}, {Armus}, {Gordon}, {Gray}, {Hinz},
  {Su}, \& {Willmer}}]{2008AJ....135.2245R}
{Rieke}, G.~H., {Blaylock}, M., {Decin}, L., {Engelbracht}, C., {Ogle}, P.,
  {Avrett}, E., {Carpenter}, J., {Cutri}, R.~M., {Armus}, L., {Gordon}, K.,
  {Gray}, R.~O., {Hinz}, J., {Su}, K., \& {Willmer}, C.~N.~A. 2008, \aj, 135,
  2245

\bibitem[{{Skemer} {et~al.}(2011){Skemer}, {Close}, {Sz{\H u}cs}, {Apai},
  {Pascucci}, \& {Biller}}]{2011ApJ...732..107S}
{Skemer}, A.~J., {Close}, L.~M., {Sz{\H u}cs}, L., {Apai}, D., {Pascucci}, I.,
  \& {Biller}, B.~A. 2011, \apj, 732, 107

\bibitem[{{Skemer} {et~al.}(2014{\natexlab{a}}){Skemer}, {Hinz}, {Esposito},
  {Skrutskie}, {Defr{\`e}re}, {Bailey}, {Leisenring}, {Apai}, {Biller},
  {Bonnefoy}, {Brandner}, {Buenzli}, {Close}, {Crepp}, {De Rosa}, {Desidera},
  {Eisner}, {Fortney}, {Henning}, {Hofmann}, {Kopytova}, {Maire}, {Males},
  {Millan-Gabet}, {Morzinski}, {Oza}, {Patience}, {Rajan}, {Rieke}, {Schertl},
  {Schlieder}, {Su}, {Vaz}, {Ward-Duong}, {Weigelt}, {Woodward}, \&
  {Zimmerman}}]{2014SPIE.9148E..0LS}
{Skemer}, A.~J., {Hinz}, P., {Esposito}, S., {Skrutskie}, M.~F., {Defr{\`e}re},
  D., {Bailey}, V., {Leisenring}, J., {Apai}, D., {Biller}, B., {Bonnefoy}, M.,
  {Brandner}, W., {Buenzli}, E., {Close}, L., {Crepp}, J., {De Rosa}, R.~J.,
  {Desidera}, S., {Eisner}, J., {Fortney}, J., {Henning}, T., {Hofmann}, K.-H.,
  {Kopytova}, T., {Maire}, A.-L., {Males}, J.~R., {Millan-Gabet}, R.,
  {Morzinski}, K., {Oza}, A., {Patience}, J., {Rajan}, A., {Rieke}, G.,
  {Schertl}, D., {Schlieder}, J., {Su}, K., {Vaz}, A., {Ward-Duong}, K.,
  {Weigelt}, G., {Woodward}, C.~E., \& {Zimmerman}, N. 2014{\natexlab{a}}, in
  Society of Photo-Optical Instrumentation Engineers (SPIE) Conference Series,
  Vol. 9148, Society of Photo-Optical Instrumentation Engineers (SPIE)
  Conference Series, 0

\bibitem[{{Skemer} {et~al.}(2015){Skemer}, {Hinz}, {Montoya}, {Skrutskie},
  {Leisenring}, {Durney}, {Woodward}, {Wilson}, {Nelson}, {Bailey}, {Defrere},
  \& {Stone}}]{2015arXiv150806290S}
{Skemer}, A.~J., {Hinz}, P., {Montoya}, M., {Skrutskie}, M.~F., {Leisenring},
  J., {Durney}, O., {Woodward}, C.~E., {Wilson}, J., {Nelson}, M., {Bailey},
  V., {Defrere}, D., \& {Stone}, J. 2015, ArXiv e-prints

\bibitem[{{Skemer} {et~al.}(2012){Skemer}, {Hinz}, {Esposito}, {Burrows},
  {Leisenring}, {Skrutskie}, {Desidera}, {Mesa}, {Arcidiacono}, {Mannucci},
  {Rodigas}, {Close}, {McCarthy}, {Kulesa}, {Agapito}, {Apai}, {Argomedo},
  {Bailey}, {Boutsia}, {Briguglio}, {Brusa}, {Busoni}, {Claudi}, {Eisner},
  {Fini}, {Follette}, {Garnavich}, {Gratton}, {Guerra}, {Hill}, {Hoffmann},
  {Jones}, {Krejny}, {Males}, {Masciadri}, {Meyer}, {Miller}, {Morzinski},
  {Nelson}, {Pinna}, {Puglisi}, {Quanz}, {Quiros-Pacheco}, {Riccardi},
  {Stefanini}, {Vaitheeswaran}, {Wilson}, \& {Xompero}}]{2012ApJ...753...14S}
{Skemer}, A.~J., {Hinz}, P.~M., {Esposito}, S., {Burrows}, A., {Leisenring},
  J., {Skrutskie}, M., {Desidera}, S., {Mesa}, D., {Arcidiacono}, C.,
  {Mannucci}, F., {Rodigas}, T.~J., {Close}, L., {McCarthy}, D., {Kulesa}, C.,
  {Agapito}, G., {Apai}, D., {Argomedo}, J., {Bailey}, V., {Boutsia}, K.,
  {Briguglio}, R., {Brusa}, G., {Busoni}, L., {Claudi}, R., {Eisner}, J.,
  {Fini}, L., {Follette}, K.~B., {Garnavich}, P., {Gratton}, R., {Guerra},
  J.~C., {Hill}, J.~M., {Hoffmann}, W.~F., {Jones}, T., {Krejny}, M., {Males},
  J., {Masciadri}, E., {Meyer}, M.~R., {Miller}, D.~L., {Morzinski}, K.,
  {Nelson}, M., {Pinna}, E., {Puglisi}, A., {Quanz}, S.~P., {Quiros-Pacheco},
  F., {Riccardi}, A., {Stefanini}, P., {Vaitheeswaran}, V., {Wilson}, J.~C., \&
  {Xompero}, M. 2012, \apj, 753, 14

\bibitem[{{Skemer} {et~al.}(2014{\natexlab{b}}){Skemer}, {Marley}, {Hinz},
  {Morzinski}, {Skrutskie}, {Leisenring}, {Close}, {Saumon}, {Bailey},
  {Briguglio}, {Defrere}, {Esposito}, {Follette}, {Hill}, {Males}, {Puglisi},
  {Rodigas}, \& {Xompero}}]{2014ApJ...792...17S}
{Skemer}, A.~J., {Marley}, M.~S., {Hinz}, P.~M., {Morzinski}, K.~M.,
  {Skrutskie}, M.~F., {Leisenring}, J.~M., {Close}, L.~M., {Saumon}, D.,
  {Bailey}, V.~P., {Briguglio}, R., {Defrere}, D., {Esposito}, S., {Follette},
  K.~B., {Hill}, J.~M., {Males}, J.~R., {Puglisi}, A., {Rodigas}, T.~J., \&
  {Xompero}, M. 2014{\natexlab{b}}, \apj, 792, 17

\bibitem[{{Skrutskie} {et~al.}(2010){Skrutskie}, {Jones}, {Hinz}, {Garnavich},
  {Wilson}, {Nelson}, {Solheid}, {Durney}, {Hoffmann}, {Vaitheeswaran},
  {McMahon}, {Leisenring}, \& {Wong}}]{2010SPIE.7735E.118S}
{Skrutskie}, M.~F., {Jones}, T., {Hinz}, P., {Garnavich}, P., {Wilson}, J.,
  {Nelson}, M., {Solheid}, E., {Durney}, O., {Hoffmann}, W., {Vaitheeswaran},
  V., {McMahon}, T., {Leisenring}, J., \& {Wong}, A. 2010, in Society of
  Photo-Optical Instrumentation Engineers (SPIE) Conference Series, Vol. 7735,
  Society of Photo-Optical Instrumentation Engineers (SPIE) Conference Series

\bibitem[{{Skrutskie} {et~al.}(2014){Skrutskie}, {Wilson}, {Nelson}, {Hinz},
  {Skemer}, \& {Leisenring}}]{2014DPS....4641815S}
{Skrutskie}, M.~F., {Wilson}, J., {Nelson}, M., {Hinz}, P., {Skemer}, A., \&
  {Leisenring}, J. 2014, in AAS/Division for Planetary Sciences Meeting
  Abstracts, Vol.~46, AAS/Division for Planetary Sciences Meeting Abstracts,
  418.15

\bibitem[{{Soderblom}(2010)}]{2010IAUS..268..359S}
{Soderblom}, D.~R. 2010, in IAU Symposium, Vol. 268, IAU Symposium, ed.
  C.~{Charbonnel}, M.~{Tosi}, F.~{Primas}, \& C.~{Chiappini}, 359--360

\bibitem[{{Soummer} {et~al.}(2012){Soummer}, {Pueyo}, \&
  {Larkin}}]{2012ApJ...755L..28S}
{Soummer}, R., {Pueyo}, L., \& {Larkin}, J. 2012, \apjl, 755, L28

\bibitem[{{Takeda}(2007)}]{2007PASJ...59..335T}
{Takeda}, Y. 2007, \pasj, 59, 335

\bibitem[{{Tamura}(2009)}]{2009AIPC.1158...11T}
{Tamura}, M. 2009, in American Institute of Physics Conference Series, Vol.
  1158, American Institute of Physics Conference Series, ed. T.~{Usuda},
  M.~{Tamura}, \& M.~{Ishii}, 11--16

\bibitem[{{Thalmann} {et~al.}(2009){Thalmann}, {Carson}, {Janson}, {Goto},
  {McElwain}, {Egner}, {Feldt}, {Hashimoto}, {Hayano}, {Henning}, {Hodapp},
  {Kandori}, {Klahr}, {Kudo}, {Kusakabe}, {Mordasini}, {Morino}, {Suto},
  {Suzuki}, \& {Tamura}}]{2009ApJ...707L.123T}
{Thalmann}, C., {Carson}, J., {Janson}, M., {Goto}, M., {McElwain}, M.,
  {Egner}, S., {Feldt}, M., {Hashimoto}, J., {Hayano}, Y., {Henning}, T.,
  {Hodapp}, K.~W., {Kandori}, R., {Klahr}, H., {Kudo}, T., {Kusakabe}, N.,
  {Mordasini}, C., {Morino}, J.-I., {Suto}, H., {Suzuki}, R., \& {Tamura}, M.
  2009, \apjl, 707, L123

\bibitem[{{Valenti} \& {Fischer}(2005)}]{2005ApJS..159..141V}
{Valenti}, J.~A. \& {Fischer}, D.~A. 2005, \apjs, 159, 141

\bibitem[{{van Leeuwen}(2007)}]{2007AA...474..653V}
{van Leeuwen}, F. 2007, \aap, 474, 653

\bibitem[{{Visscher}(2012)}]{2012ApJ...757....5V}
{Visscher}, C. 2012, \apj, 757, 5

\bibitem[{{Visscher} {et~al.}(2006){Visscher}, {Lodders}, \&
  {Fegley}}]{2006ApJ...648.1181V}
{Visscher}, C., {Lodders}, K., \& {Fegley}, Jr., B. 2006, \apj, 648, 1181

\bibitem[{{Visscher} {et~al.}(2010){Visscher}, {Lodders}, \&
  {Fegley}}]{2010ApJ...716.1060V}
---. 2010, \apj, 716, 1060

\bibitem[{{Yurchenko} {et~al.}(2014){Yurchenko}, {Tennyson}, {Bailey},
  {Hollis}, \& {Tinetti}}]{2014PNAS..111.9379Y}
{Yurchenko}, S.~N., {Tennyson}, J., {Bailey}, J., {Hollis}, M.~D.~J., \&
  {Tinetti}, G. 2014, Proceedings of the National Academy of Science, 111, 9379

\bibitem[{{Zahnle} \& {Marley}(2014)}]{2014ApJ...797...41Z}
{Zahnle}, K.~J. \& {Marley}, M.~S. 2014, \apj, 797, 41

\end{thebibliography}

\end{document}